\documentclass[a4paper, 11pt]{article}
\usepackage[margin=2.5cm]{geometry}

\usepackage[T1]{fontenc}
\usepackage[utf8]{inputenc}
\usepackage{mdframed}
\usepackage{xspace}
\usepackage{ifthen}

\newboolean{showcomments}
\setboolean{showcomments}{false}

\PassOptionsToPackage{hyphens}{url}
\usepackage[
  colorlinks=true,
  urlcolor=black,
  linkcolor=blue,
  citecolor=blue
]{hyperref}

\usepackage[
  backend=biber,
  isbn=false,
  url=true,
  doi=false,
  url=false,
  mincrossrefs=100,
  maxnames=8,
  bibencoding=utf8
]{biblatex}

\AtEveryBibitem{\clearname{editor}}
\AtEveryBibitem{%
  \ifentrytype{article}{%
    \clearfield{pages}%
  }{%
  }%
}
\AtEveryBibitem{%
  \ifentrytype{inproceedings}{
    \clearfield{pages}
    \clearfield{number}
    \clearfield{location}
  }{%
  }%
}
\AtEveryBibitem{%
  \ifentrytype{inproceedings}{
\clearname{organization}
  }{%
  }%
}

\usepackage{dblfloatfix}
\usepackage{multirow}
\usepackage{array}
\usepackage{ragged2e}
\usepackage{tabularx}

\usepackage{array,booktabs}
\newcolumntype{C}[1]{>{\centering\arraybackslash}p{#1}}

\ifthenelse{\boolean{showcomments}}{
\newcommand\red[1]{{\color{red} #1}}
\newcommand\blue[1]{{\color{blue} #1}}
\newcommand\orange[1]{{\color{orange} #1}}
\newcommand\green[1]{{\color{green} #1}}
}{
\newcommand\red[1]{}
\newcommand\blue[1]{}
\newcommand\orange[1]{}
\newcommand\green[1]{}
}

\usepackage{subfloat}
\usepackage{url}
\usepackage[table]{xcolor}
\usepackage{amssymb}
\usepackage{amsmath}
\usepackage{amsthm}
\usepackage{mathtools}
\usepackage{amsfonts}

\usepackage{footnote}
\usepackage{varwidth}
\usepackage{multirow}
\usepackage{pgfplots}

\usepackage{breqn}

\usepackage{algorithm}
\usepackage{algorithmic}
\usepackage{listings}

\usepackage{tikz}
\usetikzlibrary{arrows,calc}
\usetikzlibrary{automata,positioning,decorations.pathreplacing}
\usetikzlibrary{plotmarks}
\usetikzlibrary{snakes}
\usetikzlibrary{shadows}

\usepackage{threeparttable,booktabs}
\usepackage[normalem]{ulem}
\usepackage{setspace}
\usetikzlibrary{positioning}

\usepackage{enumerate}

\definecolor{orange}{RGB}{225,128,0}
\definecolor{brown}{RGB}{225,128,128}
\xdefinecolor{DarkGreen}{cmyk}{0.8,0,0.8,0.3}
\definecolor{rgviolet}{RGB}{119,51,255}
\definecolor{redviolet}{RGB}{255,0,85}
\definecolor{jdbrown}{RGB}{77,38,0}

\graphicspath{{./}}
\DeclareGraphicsExtensions{.pdf,.jpeg,.png,.jpg}

\usepackage{array}
\newcolumntype{P}[1]{>{\centering\arraybackslash}p{#1}}

\ifthenelse{\boolean{showcomments}}{
\newcommand{\pjv}[1]{\fbox{\bfseries\sffamily\scriptsize
    PV}~$\blacktriangleright$\textcolor{redviolet}{#1}$\blacktriangleleft$}
\newcommand{\jd}[1]{\fbox{\bfseries\sffamily\scriptsize
    JD}~$\blacktriangleright$\textcolor{jdbrown}{#1}$\blacktriangleleft$}
\newcommand{\are}[1]{\fbox{\bfseries\sffamily\scriptsize
    AE}~$\blacktriangleright$\textcolor{blue}{#1}$\blacktriangleleft$}
\newcommand{\mv}[1]{\fbox{\bfseries\sffamily\scriptsize
    MV}~$\blacktriangleright$\textcolor{DarkGreen}{#1}$\blacktriangleleft$}
\newcommand{\rg}[1]{\fbox{\bfseries\sffamily\scriptsize
    RG}~$\blacktriangleright$\textcolor{rgviolet}{#1}$\blacktriangleleft$}  
\newcommand{\commint}[1]{\fbox{\bfseries\sffamily\scriptsize
    COMM-CRITIX-ONLY}~$\blacktriangleright$\textcolor{red}{#1}$\blacktriangleleft$}
\newcommand{\comm}[1]{\fbox{\bfseries\sffamily\scriptsize
    COMM}~$\blacktriangleright$\textcolor{orange}{#1}$\blacktriangleleft$}    
\newcommand{\del}[1]{\fbox{\bfseries\sffamily\scriptsize
    DEL!}~$\blacktriangleright$\sout{#1}$\blacktriangleleft$}
\newcommand{\add}[1]{\fbox{\bfseries\sffamily\scriptsize
    ADD?:}~$\blacktriangleright$\textcolor{DarkGreen}{#1}$\blacktriangleleft$}

}{
  \newcommand{\pjv}[1]{}
  \newcommand{\jd}[1]{}
  \newcommand{\are}[1]{}
  \newcommand{\mv}[1]{}
  \newcommand{\rg}[1]{}
  \newcommand{\commint}[1]{}
  \newcommand{\comm}[1]{}
  \newcommand{\del}[1]{}
  \newcommand{\add}[1]{#1}
  
}

\newcommand{\FePCES}{\emph{EpiProtect}\xspace}
\newcommand{\FePCESnm}{Federation for Epidemic Surges Protection\xspace}

\newcommand{\PEAnm}{PriLok Entrusted Authorities\xspace}
\newcommand{\PEA}{\emph{PEA}\xspace} 

\newcommand{\APEnm}{PriLok-Associated Entities\xspace}
\newcommand{\APE}{\emph{ASE}\xspace} 

\newcommand{\EPISnm}{Edge PriLok Information Switches\xspace}
\newcommand{\EPIS}{\emph{Edge PIS}\xspace} 

\graphicspath{./}
\DeclareGraphicsExtensions{.pdf,.jpeg,.png,.jpg}

\begin{document}
\title{PriLok: \\ Citizen-protecting distributed epidemic tracing \\{\Large PRELIMINARY DESIGN V.1.1}}


\author{
   		Paulo Esteves-Verissimo, 
			J\'er\'emie Decouchant, 
			Marcus V\"olp, 
			Alireza Esfahani, 
			Rafal Graczyk\\
    SnT - Interdisciplinary Centre for Security,
    Reliability and Trust, University of Luxembourg\\
  Email: 
  \url{paulo.verissimo@uni.lu}, \url{jeremie.decouchant@uni.lu},\\
  \url{marcus.voelp@uni.lu}, \url{alireza.esfahani@uni.lu}, \url{rafal.graczyk@uni.lu}
}
  
\date{May 12, 2020. Revised on June 1, 2020.}
  
\maketitle

\thispagestyle{plain} \pagestyle{plain}

\begin{abstract}
  Contact tracing is an important instrument for national health
  services to fight epidemics. As part of the COVID-19 situation, many
  proposals have been made for scaling up contact tracing capacities
  with the help of smartphone applications, an important but highly
  critical endeavour due to the privacy risks involved in such
  solutions. Extending our previously expressed concern, we clearly
  articulate in this article, the functional and non-functional
  requirements that any solution has to meet, when striving to serve,
  not mere collections of individuals, but the whole of a nation, as
  required in face of such potentially dangerous epidemics.  We 
  present a critical information infrastructure, PriLok, a fully-open
  preliminary architecture proposal and design draft for privacy-preserving digital contact tracing, which we believe can be constructed in a
  way to fulfil the former requirements.
Our architecture leverages the existing regulated mobile communication
infrastructure and builds upon the concept of ``checks and balances'',
requiring a majority of independent players to agree to effect any
operation on it, thus preventing abuse of the highly sensitive
information that must be collected and processed for efficient contact
tracing. This is technically enforced with a largely decentralised layout and
highly resilient state-of-the-art technology, which we explain in the
paper, finishing by giving a security, dependability and resilience
analysis, showing how it meets the defined requirements, even while the
infrastructure is under attack.
\end{abstract}






\mv{
  Links from Paulo's Linkedin Article:
  \begin{itemize}
  \item tracing infection chains: \url{https://science.sciencemag.org/content/sci/early/2020/03/30/science.abb6936.full.pdf}
  \item hammer + dance: \url{https://medium.com/@tomaspueyo/coronavirus-the-hammer-and-the-dance-be9337092b56}
  \item needed for future \url{https://www.newyorker.com/news/daily-comment/the-pandemic-isnt-a-black-swan-but-a-portent-of-a-more-fragile-global-system}
  \item critique papers \url{https://eprint.iacr.org/2020/399.pdf}, \url{https://risques-tracage.fr/docs/risques-tracage.pdf}, \url{https://www.schneier.com/blog/archives/2020/04/contact_tracing.html},
  \item CT trade-offs; truth in middle: \url{https://www.lightbluetouchpaper.org/2020/04/12/contact-tracing-in-the-real-world/}, \url{https://eprint.iacr.org/2020/399.pdf}, \url{https://nadim.computer/posts/2020-04-17-pepppt.html}
  \end{itemize}
}

\section{Introduction}
\label{sec:intro}

In an earlier text published on LinkedIn\footnote{\url{https://www.linkedin.com/pulse/citizen-protecting-proximity-tracing-covid-19-between-paulo}}, we justified the
reasons behind our belief that a national infrastructure is indispensable to attend to the needs of nations in the presence of threats posed by modern pathological agents, of which COVID-19 is but an example of the future and according to specialists, perhaps a mild one.
%
%
Contact tracing (CT) is the systematic identification of
potentially infected individuals by tracing and testing those that had been in contact with a known infected person and where a transmission of the
virus may have happened. It has been an effective measure to confine the
COVID-19 outbreak in the early phase after 2020 New Year, but ceased to be
effective the moment NHS tracing capacities got exhausted. Aside from
COVID-19, the effectiveness of CT has been demonstrated in many
outbreaks~\cite{eames2003contact,swaan2011timeliness,levine2016development,roadmapPandemicResilience}.  For example, WHO
reports the essential role of CT in controlling Ebola outbreaks in
Africa\footnote{\url{https://www.who.int/csr/disease/ebola/training/contact-tracing/en/}}.
The goal of digital contact tracing\footnote{\url{https://science.sciencemag.org/content/sci/early/2020/03/30/science.abb6936.full.pdf}} is to automate CT and as such increase NHS tracing
capacities by several orders of magnitude to extend the time when CT
remains effective in chasing exponentially growing infection
rates.
However, such a proposal should preserve fundamental needs and goals, some of
which hard to reconcile, such as efficiency and effectiveness, as well as
coverage, fairness and privacy for population.

Surprisingly or not, most of the recent debate has been centred around
cryptographic aspects. However, we believe this to be a \emph{distributed critical
information systems} problem at its centre. Only by treating it with
the relevant body of knowledge will we reach the goals. By this, we
mean the right combination of s.o.t.a. ICT technologies (distributed
algorithms, fault and intrusion tolerance, networking and cloud
technology, cryptography), guided by requirements from the several
societal sectors, not only national health services and
epidemiologists, but also economists, for example.

We propose the architecture of \textbf{PriLok: Citizen-protecting distributed
epidemic tracing}, a critical information infrastructure (CII). The values we wish to
safeguard in the design we make public, are:
\textbf{
\begin{itemize}
	\item Maximizing nation-wide coverage of people and territory.
             \item Enabling controlled risk and high effectiveness decision making through whole epidemic life cycles.
	\item Transparent protection of citizens' rights, not just privacy, but also inclusiveness and fairness.
	\item Resilience against data- and system-based social and technical threats.
	\item Preservation of digital sovereignty.
	\item Protection of economy by precise and selective throttling (confinement and deconfinement).	
\end{itemize}
}

In a nutshell, PriLok should be oriented at the protection of populations,
cities, countries, trans-border regions, in the face of epidemics.
The objectives outlined above imply the participation of a plurality of
stakeholders, and for effectiveness, should leverage on existing CIIs, such as
the NHS systems and the Telco networks, both of which regulated sectors.
This preliminary proposal attempts at giving guidance to architects and
designers of infrastructures, about design avenues for devising a
citizen-protecting distributed epidemic tracing critical information
infrastructure.

The PriLok architecture is \emph{logically-centralised in concept, but
technically decentralised in implementation}, following best practices
in distributed and resilient critical information systems design.
Note that CIIs of this kind are logically centralised in nature, given
their mission. However, in distributed systems logical centralisation
does not necessarily imply monopolist trust models, or physical
centralisation. Nor decentralisation or peer-to-peer prevent abuses
per se. Both misconceptions have been part of recent
debates\footnote{\url{https://www.lightbluetouchpaper.org/2020/04/12/contact-tracing-in-the-real-world/}}\footnote{\url{https://eprint.iacr.org/2020/399.pdf}}\footnote{\url{https://nadim.computer/posts/2020-04-17-pepppt.html}}.

PriLok uses geographical decentralisation to reduce the baseline
threat plane, both at the periphery and at the core. Its management
trust model does not follow a centralised, monopolist philosophy, but
a consensual one, where abuse is technically prevented since no
operation can be executed by single or minority groups of entities,
and all critical operations require intervention of a quorum of the
(independent) entrusted entities ("checks and balances").
The core facilities themselves are also largely decentralised, distributed
and/or replicated at the entrusted entities sites. However, 
this PriLok network of components establishes
perimeter isolation from the legacy systems, with very clear entry/exit
points. This isolation is strengthened with defence-in-depth mechanisms implementing a high
degree of fault and intrusion tolerance. The resulting threat plane reduction in face
of external and internal attacks or faults is a key aspect, to achieve
resilience in general, and privacy in particular, despite handling critical data.

Unlike some recently published approaches (e.g., exclusively based on
Bluetooth), we favour technologies that promote incremental inclusion
of all population strata --- economic, literacy or age.  Given the
significant percentage of population estimated not to own a
smartphone and/or not being tech-savvy, we see currently
no alternative to the mobile communication system as a baseline.

We aim as well at protecting digital sovereignty, avoiding as much as
possible solutions that open considerable threat planes like those
affecting phone-to-phone attacks or generating inconsiderate
dependence on phone/OS vendors, which might for example cause massive
leakage of national critical data to unidentified threat actors.

Finally, we have learned from past and present epidemics that even
small delays or imprecisions in decision making, can have damaging effects,
on health if by default, or on economy if by excess.  For example,
super infectors or infection hotspots require immediate and precise
identification and isolation, especially in the beginning, or in a
deconfinement phase when people relax. 
On the other hand, policies seeking herd immunity, which seem a very logical
approach but have dramatically failed during COVID-19, might have been
successful had such infrastructures as we propose here been present
from day one, following and predicting the situation accurately and
timely.
Likewise, indiscriminate closing of the economy has the devastating
effects that we have been observing in COVID-19 times.

In consequence, unlike some recently published approaches (e.g.,
decentralised and voluntarist) we consider that only an approach based
on a logically centralised global view of the epidemic evolution can
provide the accuracy, near real-time situational awareness and
predictive power, required for controlled risk and high effectiveness
reactions. Such a CII will, in our opinion, significantly mitigate the
risks of upcoming epidemics and possibly prevent pandemics, by enabling
precise measures of the national health systems, as well as the
economy-saving throttling of the society activity.

\mv{ Logically centralized Identification of cluster members is
  particularly useful for diseases with some, but not significant
  superspreading characteristics (e.g. in the order of \kappa = 0.45)
  which require immediate Identification and isolation of all members
  of a potential cluster in which such an event occurs.  Sars cov 2
  seems to be exactly auch a thing and we have No Time to loose
  isolating Potential Clusters when we want to benefit from
  superspreading. I'll dig out the refs when the transcript of todays
  Podcast by Christian Drosten is Public.  }

In a nutshell, our proposal attempts at \textbf{striking a balance between
securing health, protecting privacy and safeguarding the economy.}

\section{Requirements specification}
\label{sec:req-spec}

To be clear about the objectives and trade-offs of PriLok, we discuss below
all the desirable requirements that we believe should be met by an
infrastructure of this kind, and the rationale for meeting them.

\subsection{Desirable objectives and implied requirements}
\label{ssec:obj-impl-req}

We list the almost indispensable functional objectives that should be reached by any nation-level critical infrastructure doing digital contact tracing (CT) (R1-R6):
\begin{enumerate}[{R}1]
	\item Be epidemic-agnostic: able to act on any epidemic, even the unexpected,
				in near real-time.
	\item Help find the highest possible rate of infected individuals in near
				real-time.
	\item Help find reasonably complete and accurate potential infection chains
				in near real-time.
	\item Alert, monitor, confine, and trace potentially infected individuals
				in near real-time.
	\item Diagnose country/region/community epidemic dynamics in near real-time
				(map basic infection evolution numbers; locate and map infection
				hotspots and trajectories; detect super infectors and/or lone wolves;
				predict collections of asymptomatic individuals; discern between
				external and communal infection paths).
	\item Incorporate lessons and feedback from first epidemic outbreaks and adapt further actions during individual
				re-infections and epidemic recurrences, in near real-time.
\end{enumerate}

\vspace{.2cm}
\noindent
Additionally, the following non-functional objectives should be met (R7-R10):
\begin{enumerate}[{R}1]
	\setcounter{enumi}{6}
	\item Guarantees of protecting citizens' fundamental rights (such as
				transparency, privacy and equality) in compliance with the law.
	\item Resilience to manipulation and forging, fake-news, gossip, panic,
				denial of service.
	\item Sustained real-time capability under overload, to maintain situational
				analysis and reaction capacity (infection roadblocks; sanitary fences
				around hotspots; group quarantines; and later, precise selective
				re-opening).
	\item Smoothly incremental accuracy and recall of proximity
          event determination, from an inclusive though possibly
          coarse \add{sovereign} nation-wide baseline technology
          level, to finer levels attainable by s.o.t.a.  technology (not
          only but including 5G).
\end{enumerate}

\mv{We have digital sovereignity as a thread, but not as an objective; added it to R10}

\subsection{Rationale}
\label{ssec:rationale}

If those requirements are met, we are bound to have a CII (critical information
infrastructure) that really serves a nation and its individuals, in the possibly
hard times to come in the next years\footnote{\url{https://www.newyorker.com/news/daily-comment/the-pandemic-isnt-a-black-swan-but-a-portent-of-a-more-fragile-global-system}}. Furthermore, their correct implementation
guarantees that the 7 fundamental principles of the GDPR~\cite{eu:gdpr} are followed:
lawfulness, fairness and transparency; purpose limitation; data minimisation;
accuracy; storage limitation; integrity and confidentiality; accountability.


\emph{The possible criticality and magnitude of future epidemic surges advises
that nations be prepared: instead of reactive, be proactive. Moreover, the time
is now, and not during the next epidemic surge.}

It took two large tsunamis for practically all coastal countries to
set up CIIs that are permanent, working as tsunami alert, follow-up and
prediction systems. Countries should probably realize sooner than later
that it is time to have an epidemic alert and evolution
prediction system in permanence.

This is why we believe this is a task for the \emph{\textbf{state}}, as one stakeholder
of the nation. It should have the important responsibility (political as well as economic)
of its implementation and operation, relying on other stakeholders (regulated
companies, regulators, public associations, for example).

However, the \emph{\textbf{people}}, individuals or collections thereof
(who ‘are' the nation) have a right to enjoy the CII on an equal basis
(regardless of their technical literacy), release PII (personally identifiable
information) lawfully, only on a need basis, by the principles of storage
limitation and data minimisation, and having transparent access to its design
and operation auditing.

\emph{It would be excellent if no involvement of PII would be needed, given the
criticality, but such an infrastructure, if it is to protect the nation,
it must get to the nation.}

There are currently a number of proposals for digital contact tracing, including DP3T\footnote{https://github.com/DP-3T/documents}, TraceTogether, ROBERT\footnote{https://github.com/ROBERT-proximity-tracing/documents}, TCN\footnote{https://tcn-coalition.org/}, NTK\footnote{https://github.com/pepp-pt}, Canetti-Trachtenberg-Varia~\cite{canetti2020anonymous}, the Apple/Google's joint initiative for Bluetooth distance measurement in iOS/Android\footnote{https://www.apple.com/covid19/contacttracing/}, Pronto-C2~\cite{prontoc2}, PACT-WEST~\cite{chan2020pact}, PACT-EAST~\footnote{https://pact.mit.edu/wp-content/uploads/2020/04/The-PACT-protocol-specification-ver-0.1.pdf}, Reichert-Brack-Scheuermann~\cite{reichert2020privacy}, etc., which we
do not wish to criticize, since all contributions are not too many in
these critical times. We believe nevertheless that a good test of their fitness
for the purpose would be for the authors showing that they pass the Litmus test
of meeting the requirements R1-10 above. Some proposals, however focused,
present very elegant algorithmic solutions to parts of the big picture addressed
by PriLok. We do not exclude the possibility of considering their
contribution within the skeleton provided by PriLok.

In this sense, we believe that approaches \emph{\textbf{peer-to-peer managed}} 
(actually or pseudo-decentralised), \emph{\textbf{and voluntarist}} (totally or
mostly based on word-of-mouth gossip), will work to a certain point, but will
miss some important objectives of the list above, not least, the equality of
access and coverage of population, and the capacity for global (nation-wide)
and timely reasoning.
However, approaches \emph{\textbf{centrally managed}}  (single entity), 
and \emph{\textbf{top-down controlled}} (totally or mostly based on
the \guillemotleft Trust me because I tell you
to!\guillemotright \space principle), and as such \emph{\textbf{opaque}}, will work as
well, but miss another set of equally important objectives listed, not
least, by losing confidence of the people in terms of privacy, and
perpetuating a state of surveillance.

\emph{It is our opinion that the risks impending on the PII can be significantly
mitigated, with an adequate mix of the right social/political management
framework and state-of-the-art technical measures to safeguard the information.
This being achieved, the benefits (R1-10) will largely outweigh the risks.}

An infrastructure such as we envisage, albeit supported by the state,
must not be built or managed in a fully centralised way. It should
instead be managed in concertation through \emph{\textbf{consensual
    actions by several powers exerting mutual control}} (“checks and
balances”), in respect for the PII it will store and
process.  Correctness of these consensual actions must of course be technically
enforced by robust technologies, such as protocols of the BFT class
(Byzantine Fault Tolerance), playing together with multiparty
cryptography protocols. These technologies, albeit sophisticated, have
today a high technology readiness level (TRL), spawned by its
increasing use in a number of real world applications, notably the
Fintech/Blockchain area.

Furthermore, the infrastructure should be dormant (locked and largely empty
of information) most of the time, only to be activated in times of
need, by multiparty decisions; PII information collected should be
disposed of immediately it is no longer needed; PII information at
rest during active periods should be protected with strong multiparty
cryptography, and so forth. In consequence, such an infrastructure
must be designed and implemented using the best technical practices
available to ensure all these objectives.

For this to be done without large impact on the efficiency, the decision and
operation processes should be streamlined and based on IT-supported workflows,
but attested and certified continuously (e.g., by indelible logging apparatus
and/or blockchain supported ledgers). Ex-ante and ex-post auditing should be put
in place, effected by an independent regulation body. Citizens should as well
have transparent access to the modus operandi and the results of the regulation actions.

\section{Preliminary Architecture and Draft Design}
\label{ssec:arch-draft-design}

We present a fully-open Preliminary Architecture Proposal and Draft Design of
PriLok. Our purpose is not to give a fully-fledged design, but rather to give
guidance to architects and designers of such infrastructures, should these ideas
merit the support of the main stakeholders in a nation, certainly the state and
the citizens. As such, we do not intend to go into too much further detail in
the sections below, beyond giving the outline and skeleton of protocols and
mechanisms, showing that the main architectural, data model and algorithmic
design options meet the requirements R1-R10. The design is also open enough
that, within the margin we leave for the technical options, different nations
may strike different balances between securing health, protecting privacy and
safeguarding the economy.

\subsection{Introduction}

Generically, the PriLok infrastructure is implemented and controlled by a
“\FePCESnm - \FePCES”. In the context of this paper, 
\FePCES is the designation of the necessary  coalition of interest formed by entities of the state --- such
as relevant government ministries, National Health Service entities like centres
for disease control and hospitals, Justice, an independent Regulation body for
the CII --- and regulated companies, regulators, research and technology
institutes and universities, public associations, for example.  The PriLok
infrastructure, albeit supported by the state, is managed not in a fully
centralised way, but in concertation, by several powers exerting mutual control
(“checks and balances”), in respect for the PII it will store and process.  In
essence, the \PEAnm (\PEA)
is a subset of
the entities listed above, whose number and quality/role will depend on specific
countries' culture and legal systems. \PEA members are those that can collectively issue
authorisations for the manipulation of PriLok. As seen below, all such
operations must be vetted by a quorum of the \PEA. Other entities such as listed
above will be
\APEnm (\APE).

\subsection{Architecture components}

\begin{figure*}[ht]
  \includegraphics[width=\textwidth]{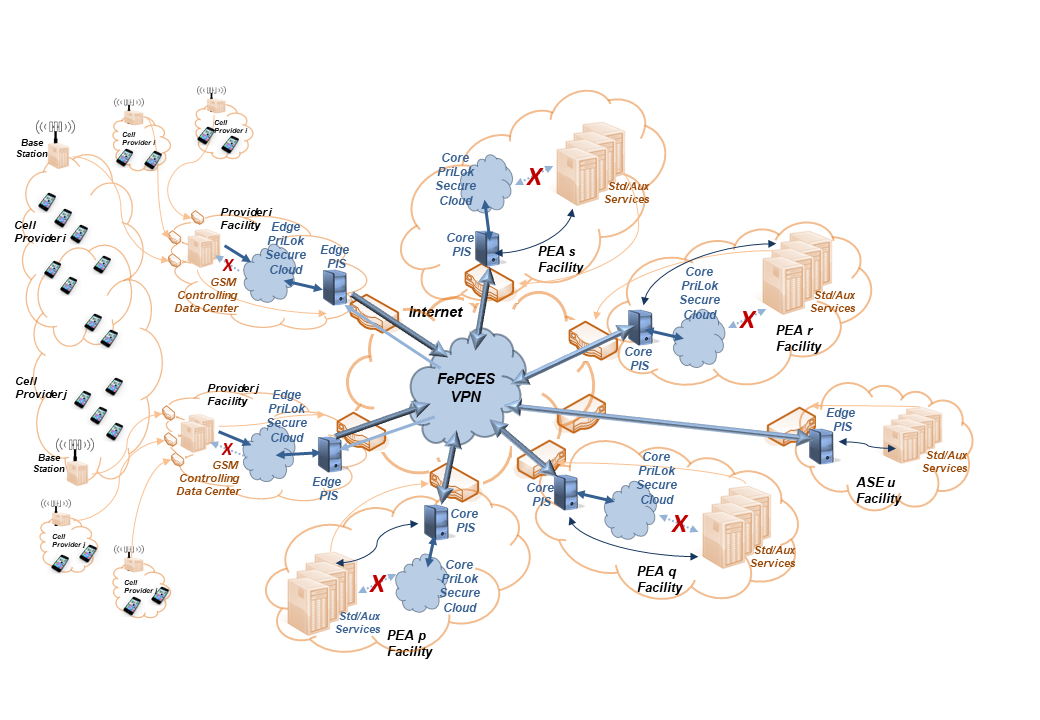}
  \caption{PriLok architecture}
  \label{fig:architecture}
\end{figure*}

Figure~\ref{fig:architecture} gives an overview of the
  architecture. It is worth noting that, following a successful
  concept in previous research on critical information
  infrastructures, the technology required is already available (e.g.,
  but not exclusively from the EU projects CRUTIAL\footnote{https://cordis.europa.eu/project/id/848109}, MASSIF\footnote{https://cordis.europa.eu/project/id/257475}, BBC\footnote{https://cordis.europa.eu/project/id/317871}, SEGRID\footnote{https://cordis.europa.eu/project/id/607109}).
PriLok attempts at leveraging (rather than replacing or duplicating) existing
CIIs, in this case the legacy Telco and Public Administration infrastructures in
general, and the Mobile Communication system in particular. As such, as seen in
the Figure, whilst the existing legacy systems are represented in brown, PriLok
is laid out as an overlay architecture over them, represented in blue.
Furthermore, to ease integration and cause minimal disturbance, PriLok
components are highly modular and self-contained (information switches, cloud
subsystems). This perimeter isolation with very clear entry/exit points
(boundaries between brown and blue in the Figure) is also key to security and
dependability. As we show ahead, it is strengthened with other defence-in-depth
mechanisms in order to attain the very high levels of resilience desired.\\

\subsubsection{\textbf{Edge realm}}

\noindent \textbf{Telco Operator and Service Provider (Provider in short) cellular network cells such as macrocells, microcells, picocells, and femtocells (existing).}

\noindent	We leverage the existence of 
	the cellular public network, since any ‘live’ mobile phone will be in contact with at least one Provider (or potentially more, in case of roaming),
	in any covered location. PriLok is set up as an overlay architecture over/aside the cellular systems, and tries to cause
	the least disturbance possible on the cellular network. However, we consider that
	PriLok, as a regulated infrastructure of public interest, may reasonably imply
	some minimal changes on the Providers, as described below.
        
	Currently, the  cellular network has a degree of variation in the implementations, according to
	Providers’ structure and xG generation. In what follows, we provide a general
	outline of a prototypical architecture, for simplicity and without loss of
	generality. In cellular networks, small cells are employed to enhance the link quality and network capacity~\cite{liu2013deploying}. 
	Several types of small cells include femtocells, picocells, microcells, and macrocells – broadly 
	increasing in size from femtocells which are the smallest, to macrocells which are the largest. 
	The network is normally organised in cells, nominally covering a
	geographical region, by sets of antennas controlled by the cell Base Station.
	Cells from the same provider, or from different providers, overlap in their
	spatial coverage.
	
\begin{figure*}[ht]
          \begin{center}
		\includegraphics[width=.8\textwidth]{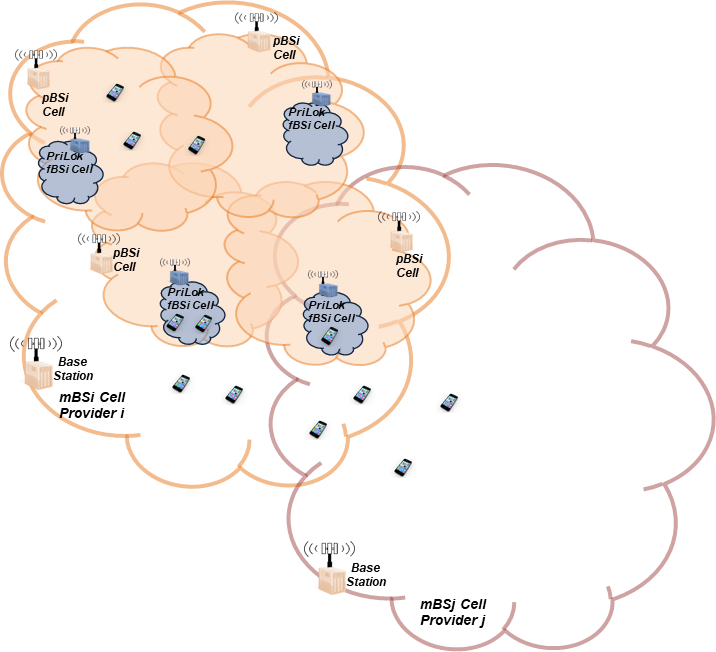}
		\caption{PriLok integration with cellular network infrastructure}
		\label{fig:integration}
          \end{center}
	\end{figure*}

	The Figure~\ref{fig:integration} suggests the current reality of the cellular (mobile
	communications) system, and the small add-ons that may be implanted by PriLok
	(fBS in blue, explained below). Macrocells (standard cells and microcells)
	implement the external (street) structure, respectively by macrocell Base
	Stations, mBS. Communication inside premises (e.g., internal parkings, theatres,
	shopping malls) is secured by additional, finer granularity, picocells, from a
	given provider, controlled respectively by picocell Base Stations, pBS. These
	are aggregated under the realm of the macrocell that subtends them, by a
	hierarchical logical structure, called paging cell. The useful ranges of the pBS
	of a same paging cell partially overlap in their spatial coverage. A phone will
	register to a cell upon arrival (e.g., through the macrocell base station mBS),
	and after that the communication enters stand-by listening mode to save energy.
	From then on it can be paged by any of the base stations in that paging cell
	(e.g., walking through a shopping mall) on a need basis (e.g. an incoming call
	or SMS).
	
	The information flow will be detailed in a later section, but a key data
	structure for the process described below is introduced now:
	\emph{Proximity Detail Record (PDR)} 
	--- containing, for each region (cell), the timestamps of contiguous
	periods of time spent in the region by a phone, the average proximity vector 
	from the centroid, plus an encoded ID of the region BS.
	
	\comm{
		average proximity vector  from the centroid, - hmm, don’t know if this is the
		best info; see also figure in PriLok information flow.pptx}
	\comm{  encoded ID of the region BS -  mentioned elsewhere, we should not
		know WHERE is the BS, ‘coz that is  absolute LOCATION, not relative
	}
	
	Technically, it is possible today that:
	\begin{enumerate}[(i)]
		\item mBS or pBS calculate the proximity of a phone to the centroid of the
		users distribution of the respective antenna set, as a relative position; 
		\item several pBS of a paging cell can periodically page a phone on
		purpose to determine that proximity 	vector  (e.g., a polar coordinate
		from the centroid) and generate a PDR; 
		\item the mBS and several pBS of a paging cell can triangulate their
		space-time readings of proximity of a same phone, in order to get a much
		more precise value of the relative position of the phone relative to the
		antenna set, and create the respective more precise PDR; 
		\item in alternative, that triangulation can be performed later, over
		independently recorded PDR registers by several BS, containing
		space-time readings of proximity of a same phone to those BS, relative
		to a similar time interval; 
		\item none of these registers need to contain absolute location information.
	\end{enumerate}
	\vspace{2mm}
	
	\comm{
		need to check one thing for later (info flow) – it should not be possible to
		know by looking at a PDR, the absolute location of the BS. Ergo, we should not
		know to which BS the coded ID corresponds. The provider knows that of course. If
		later needed for e.g. hotspot id, a request is made to the provider. So, let’s
		think of a smart way of encoding the BS ID, which of course must be in the PDR.
		Smart mostly in terms of key mgt...
	}
        
        \commint{I (Marcus) don't see why this is problematic. The PDR
          is encrypted, so a normal level of indirection helps. I.e.,
          we have a base station ID, which merely defines which
          decryption key to use. Then, $PDR$ can simply include
          $\{(\mathit{Base location})\}_{E_{\mathit{pub},BS}}$}
	
\noindent \textbf{Edge Telco Provider Cellular Controlling Data Center (existing).}\\
We denote the cellular Controlling Data Centers (DC) generically (provider
implementations vary), as the first DC on the edge of the provider network where
PDRs collected by the base station network can be concentrated and stored
systematically.\\

\noindent \textbf{Edge PriLok Secure Cloud (PSC) (in Telco Provider Cellular Controlling Data
Centers).}\\
The Edge PriLok Secure Clouds (PSC) are the PriLok-supplied subsystems
co-located in each Controlling Data Center of Providers. The PDRs are stored
encrypted in this installation as they come from the BS, by the Provider, which
has a write-only (push) interface to the Secure Cloud. After PDRs are stored in
the cloud, they can no longer be accessed by the provider.\\

\noindent \textbf{Edge PriLok Information Switches (Edge PIS) (containing the edge services and connection to the VPN).}\\
The Edge PIS are the points of contact of the edge secure clouds with the core
systems, through the \FePCES VPN (see below). They also run services that manage
the information in the secure cloud. We foresee that these data secure the
principles of data minimization and storage limitation followed by data
protection authorities and the GDPR in general, by: containing minimal
information about phones only; ibid about presence under cellular network cells, i.e.
relative location, i.e. proximity, not absolute location; being automatically
deleted after a time-to-live period to be defined, a function of the target
disease incubation time.

\subsubsection{\textbf{BASE 0 - Proximity Tracing with Cellular components}}

	Proximity Tracing with cellular components has incremental levels of precision, from
	older xG or e.g. rural areas where the useful range of mBS may be kilometres,
	through metropolitan areas where it may be a couple dozen meters or less, to
	inside premises pBS, where it can come down to a few meters. This approach shows
	a virtuous adaptation of accuracy to human density, providing a predictable rate
	of false positives grossly proportional to predicted urban density. The approach
	also promotes inclusion, since: 30\% of the population is estimated not to own a
	smartphone, and most older people are not tech-savvy. Thus, including older
	people in a system that works automatically for them, and with a predictable
	rate of false positives grossly proportional to age (and thus health risk) and
	tech-illiteracy, seems as well a virtuous trade-off for an infrastructure of
	public interest. From the viewpoint of the national interest, it is also the one
	offering better reliability, security and sovereignty conditions for a start,
	since it does not suffer from the considerable threat plane affecting
	phone-to-phone attacks, or the phone/OS vendor interference (both in GPS and in
	Bluetooth sensing).
        
	Experiments will need to be done to determine the actual levels of accuracy and
	recall of contact tracing allowed by the several technical levels of the baseline system (mBS, pBS,
	triangulation) as we have described.
        
	Some things should be noted however: (i) the problem with older generation
	equipment is expected to lie mostly with accuracy, i.e. in alerting too many
	people, rather than missing infected/infector people; (ii) if our conjecture is
	correct, this would concern the virtuous combinations mentioned above and thus
	be a good trade-off; (iii) on the other hand, accuracy will be most important in
	points likely to become infection hotspots (packed-layout restaurants and other
	commercial surfaces, bars, theatres, sports halls, PoS, etc.), where again,
	newer generation equipment is more expected; (iv) next we discuss ways to
	further improve accuracy thinking about these spots.

\subsubsection{\textbf{BASE 1 - Proximity Tracing with Cellular Enhanced components}}

	\noindent \textbf{PriLok cells.}\\
	We go further in solving this remaining problem and improving the precision and
	accuracy of contact tracing given by this baseline architecture, by selectively
	enhancing them in the most needed points (as the examples just above). We go
	down one order of magnitude in spatial range, inspired by the femtocell 
	principle in mobile networks. Femtocell is a small, low-power low-capacity base station,
	with a useful range of a few meters, typically designed to solve coverage corner
	cases, or serve homes or small businesses.
        
	The analogy stops there, and we introduce PriLok special femtocells (depicted in blue in the
	Figure \ref{fig:integration}), implemented and controlled by dummy base stations
	that we call fBS. Inside a given paging cell, there may be several fBS,
	installed in consonance with the respective Provider.
	\comm{
		ah, maybe a PriLok fBS can talk to more than one provider…, those overlapping
		in an area? That would be very interesting, and doable from a CII viewpoint…;
		otherwise, e.g. a restaurant might have to have 4-5 fBS
	}
	fBS present themselves to phones as genuine base stations of a paging cell. So
	they can force the periodical paging of a phone in the (very small) area of
	their useful range. After each ping, they do not perform mobile communication,
	which is ensured by having the phone connect to another pBS in the area with
	overlapping coverage.
	
	Technically, it is possible that mBS and pBS are software-enhanced (with few
	exceptions) so that fBS can interact with mBS and pBS nearby in a simple manner:

\begin{enumerate}[(i)]
	\item by having them calculate and store the proximity of the fBS the same way
				they do with phones, triangulating their space-time readings in order
				to get a precise value of the relative position of the fBS relative to
				the antenna set (this operation is done once per fBS set-up, since the
				fBS is not expected to move relative to the mobile system BSs, in
				principle);  
\comm{
fBS should of course be able to move for the advanced application scenarios,
where we place them inside a bus or in a train, but since both are closed rooms
anyway, we don't need proximity to neighboring regions, i.e., if an infector is
in a bus, he might have infected all that took the bus at the same time; PJV:
the fBS does not communicate, needs a BS near… so I agree buses and trains were
and are an option for us, we discussed that with the initial idea for the
PriLok box; I even have somewhere an example with a bus; but don't know exactly
how we will handle that with fBS (other than having fBS sync with passing BSs as
do phones….), however, the PCscor of all these guys inside, passing several BS
together for 30min or an hour, man, maybe we don't need an fBS…) 
}

	\item whenever this is not possible, the fBS can be georeferenced by hand
				through a GIS of the area. 
	\item by sending the related paging events of phones that enter and leave
				their range to one of the mBS or pBS (which issue a PDR with the
				respective timestamps, the average proximity vector of the fBS from the
				centroid of the issuing BS, and an encoded ID of the latter). The PDR
				thus contains a point with much higher precision than what is achieved
				even by picocells.
\end{enumerate}

\comm{
relative position of fBSs is known by the upper mBS or pBSs, but should not go
in the PDR…
}

\comm{(check/decide what is telco, what is gov)}

	\noindent \textbf{Alternative proximity tracing technologies.}\\
PriLok assumes a default baseline measurement approach based on
the cellular apparatus, for inclusion, fairness and completeness.
Then, it improves on the baseline through the above-mentioned described PriLok pseudo femtocell.

However, it welcomes integration of other approaches, for example
those working on a voluntary basis, possibly for complementing
information in specific situations and areas, e.g., GPS, Bluetooth,
Wifi or other. 

However, this must be done with care, always taking into account the non-functional objectives (R7-R10), in
particular digital sovereignty.

\subsubsection{\textbf{Virtual Private Network (VPN) realm}}

\emph{This block is essentially materialised by the protocols
  implementing the \FePCESnm (\FePCES) VPN, linking the institutions
  entrusted to manage epidemic tracing, and associated
  institutions.}
  
The VPN is supposed to interconnect all nodes of
the architecture, through Edge and Core PriLok Information Switches:
Edge PriLok Secure Clouds (PSC); Core PriLok Secure Clouds (PSC);
PriLok Complex Event Processing Engine (PCEPE); PriLok Data Vault (PDV); any \PEAnm
(\PEA) not co-located in one of the facilities listed above; and
privileged \APEnm (\APE) needing secure access.

\subsubsection{\textbf{Edge realm: associated entities}}

\noindent \textbf{\APEnm (\APE) facilities (existing).}\\
PriLok is destined to fulfill several societal objectives. As such, it is
natural that one of the needs is the secure information export to, or import
from, external entities needing to work on it. The particular information may or
not have privacy criticality.

In consequence, PAE that only need to receive or send non-critical information
will do so by standard information transfer mechanisms. PAE that need to receive
or send critical information as well MUST do so via mechanisms provided through
the \FePCESnm (\FePCES) VPN.

This will be implemented by means of a protocol to be established between the
\FePCES and the relevant PAE, and materialised through an Edge PriLok Information
Switch (Edge PIS) connected to the VPN, similar to those used in the Telco
Providers edge.

Any significant amount of critical information leaving the PDV to \APE (\APEnm),
e.g., for research purposes (such as statistical collection
and epidemics modelling), should provide strong guarantees of
anonymity and generic protection of any PII (that has in the meantime
not been made non-private, e.g., according to the laws of some
countries with regard to notifiable diseases).  N.B.- The words of
caution made about in-core workflows under more sophisticated
operations are echoed here by majority of reason, for externalisation
of information to associated entities or the public. Before allowing,
in further versions of PriLok, more aggressive release of information
without raising the risk, and additionally to what was suggested for
the improvement of the security of the in-core workflows, further
research is suggested on the investigation and verification of
algorithms allowing privacy-preserving information disclosure, for
example leveraging s.o.t.a. on
k-anonymity~\cite{sweeney2002k} and its successive refinements~\cite{machanavajjhala2007diversity, li2007t}, or differential privacy~\cite{dwork2008differential, dwork2010differential, dwork2010pan}.\\





\subsubsection{\textbf{Core realm}}

\noindent The Core realm consists of facilities containing the
storage and computing capacity to handle the PriLok operation. To be
instrumented in facilities of the \PEAnm (\PEA), as extensions of
existing installations, or created a new.\\

\noindent \textbf{Core PriLok Information Switches (Core PIS).}\\
Containing some core services and the connection to the VPN).\\

\noindent \textbf{Core PriLok Secure Clouds (PSC).}\\
In the \PEA Data Centres, containing
private storage and compute clouds supporting other components, see below).
The Core PriLok Secure Clouds (PSC) are a compound of clouds in the core
facilities of PriLok (\PEA). They offer the decentralised basis for running
distributed protocols implementing the PriLok Complex Event Processing Engine
in a distributed and parallel instantiation, and the PriLok Data Vault in a
resilient (fault and intrusion tolerant) way. Both are described below.\\

\noindent \textbf{PriLok Complex Event Processing Engine (PCEPE).}\\
PCEPE is the engine where
computations are massively run, in principle implemented in one or more Core PriLok
Secure Clouds (PSC), in \PEA Data Centres.\\

\pjv{(describe CEP, look at MASSIF, and NeXcale?; I have a survey paper from
2008, but here we want mature tech, and not too much detail, should do:
data-streaming-D2.1-SOTA)}

In PCEPE, collected events (paging information) are processed and tracing
information is extracted and stored in data vault. PCEPE has to operate on
streams of information, in near real-time and, above all, has to be implemented
as trusted computing service. PCEPE is designed as a streaming system as it has
to run continuous queries on constantly arriving input data, in order, to capture
the ever-changing locations of potential subjects. On the contrary, batch
processing would require storing of large volume of raw data and would be
inefficient for this purpose.
PCEPE architecture has to be dependable, and at the same time, scalable. Such
complex processing engines already exist as a research prototypes, i.e. Massif
project \cite{Garcia13siem, Garcia2016sieveq, Sousa2013bftsmart-tech-report},
but also reached maturity level where they have been adopted by industry and
deployed in production, including BeepBeep-3 \cite{beepbeep3}, Apache Flink and
Storm \cite{apachestorm}, SQLstream \cite{sqlstream}.\\

\begin{figure}[ht]
  \center
  \includegraphics[width=.9\textwidth]{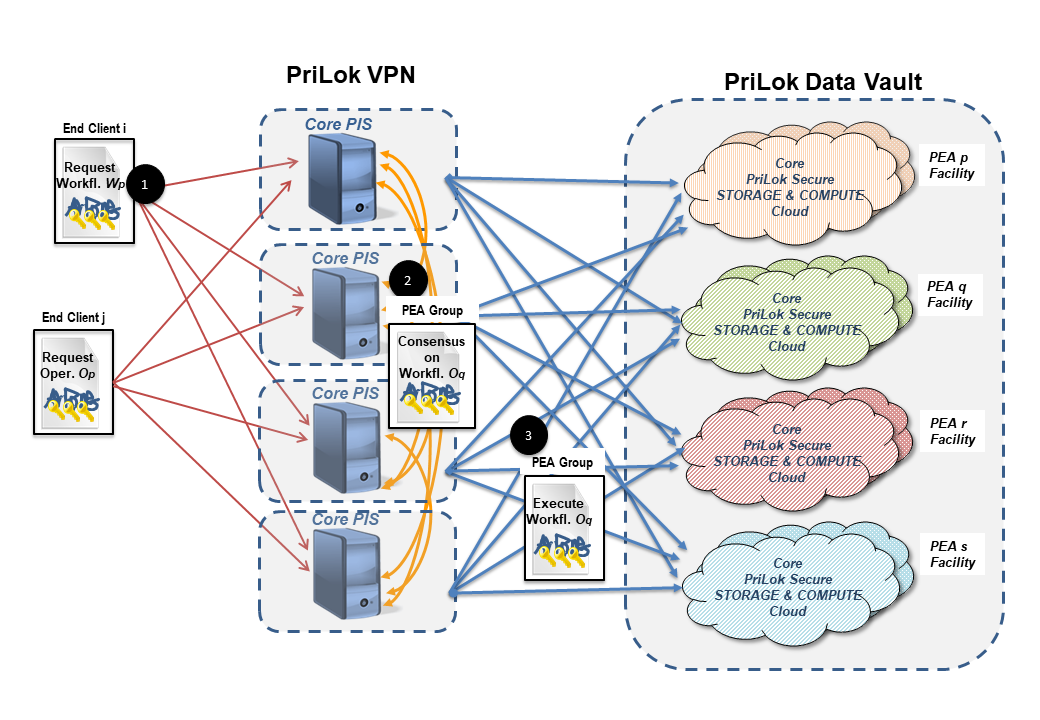}
  \caption{PriLok Data Vault: Consensual triggering of operations}
  \label{fig:checks_balances}
\end{figure}

\noindent \textbf{PriLok Data Vault  (PDV).}\\
PDV is the main data repository.  Though logically centralised, the
PriLok Data Vault (PDV) construction is NOT physically
centralised. It is distributed, as depicted in
Figure~\ref{fig:checks_balances} and as we explain below, amongst
several \PEA entity nodes, an independence that provides
decentralisation of operation and resilience to faults and
attacks. PDV is essentially a data store, in principle key-value in
its nucleus, implemented by one or several core private storage
clouds where pre-processed and post-processed data are stored.  To
reap performance benefits, the compute clouds needed to perform the
PriLok workflows are co-located in the same \PEA facilities, as shown
in the figure.  The highly secure workflows PriLok is destined to
run, are coordinated from distributed protocols running on the VPN, in
the several core PriLok information switches (Core PIS) which,
recalling Figure~\ref{fig:architecture}, isolate and connect the
PriLok components running in several facilities.

\mv{Turn into floating figure}

The PriLok VPN and Data Vault compound represented in
Figure~\ref{fig:checks_balances} is a crucial building block which
builds on a large body of knowledge on fault and intrusion tolerance
and resilience (e.g., Byzantine fault-tolerance, cloud-of-clouds
tech., multiparty cryptography, erasure coding, etc.).  
One of the central fears and a threat to be reckoned with is the
execution of sensitive operations by a single or minority groups of
entities. 
PriLok addresses these possible threat vectors by requiring consensus
for all critical operations by a quorum of independent entrusted
entities. At the level of machines invoking services at other
machines, classical solutions, such as Byzantine fault tolerant
state-machine replication protocols (e.g., PBFT~\cite{pbft},
MinBFT~\cite{veronese2011efficient}, CheapBFT~\cite{cheapbft}, but
also variants deployed in modern
blockchains~\cite{androulaki2018hyperledger,abraham2016solida,gilad2017algorand,gueta2019sbft,avalanche,eyal2016bitcoinng,kogias2016enhancing})
are readily
available. 

We detail the security and dependability aspects of the PriLok VPN
and Data Vault compound represented in
Figure~\ref{fig:checks_balances}, in the next section.

\comm{
At the personal level of humans interfacing with the
PriLok critical information infrastructure, the same principles can
be applied in the workflows that govern these
operations. Figure~\ref{fig:checks_balances} gives an example of a
laboratory assistant, just having positively identified a patient,
initiating a search of contacts, by reporting the case to an NHS
representative and possibly to a judge to confirm this operation. Each
of the three confirm the request (e.g., by digitally signing it) and
the platform acts only on properly signed requests. \add{Roles thereby
  help avoiding denial of service, by allowing multiple individuals to
  assume a role (such as the three sketched in the figure). And
  majority quorums can be used to overrule non-compliance of all
  individuals in a given role.}\mv{I'm not sure we need this
  addition.}}

\subsubsection{\textbf{Security and dependability aspects}}

\noindent \textbf{Security and dependability of Core subsystems: Policy aspects.}\\
Queries, and direct reads and writes can be made on PDV under incremental
authentication and authorisation policies established by policy makers, issued
by quorums of the \PEA entities and implemented by the technologies underlying
PDV.

As explained elsewhere, PriLok follows generically a 2q-eyes access
control policy: it is necessary that q entities amongst the n entrusted ones,
vet any transaction that modifies or extracts information from the Vault. The
size of quorum q may vary with the class of operation (also discussed
elsewhere).

Generically, depending on the operation, q corresponds for example to x+1
minimum number of unblocking shares of an (x+1, n) multiparty crypto operation,
such as a threshold signature, or the recovery of the (x+1, n) shared key
protecting PDRs in edge clouds, or other processed records in the core data
vault clouds. The quorum q may also correspond to some f+1-fault tolerant quorum
of entities needed to secure a majority vote on operations that relinquish, or
allow modification of, information in the PDV.

The workflow to gather the necessary authorisations should be apparent to the
entities involved. For example, when requested by one of the entities involved
in the activities of the Epidemic Tracing and Prediction Federation, it only
goes ahead after being authorised by enough other entrusted entities. For this
to be done without large impact on the efficiency, decision and operation
processes should be streamlined and based on IT-supported workflows (e.g.,
through some form of ERP systems workflow support).

In the example of Figure~\ref{fig:checks_balances}, such authorised
requests for single operations or workflows (1) (gathering the
necessary number and qualities of signatures) are arriving at the
PriLok interface, broadcast to all core PIS. The BFT protocols in the
PIS run in order to reach a consensus (2). Each PIS resides in a facility
that is managed by and represents an independent stakeholder of the
system, as we have discussed before. That is, even in the presence of
$f$ faulty players or attackers, in the end there is at least a majority number
of correct players agreeing on what the workflow should be, and thus ending-up deciding to 
execute the correct workflow (3).

Now, the workflow, as depicted in the Figure, combines access to the
data at rest in the storage clouds, with the computational elements in
the compute clouds, for example, the PCEPE. The workflow is triggered
by the BFT protocols in the PIS, ensuring that it is correctly
implemented, and maintaining the security properties desired of the
Vault data, namely privacy.  Again, no data can be extracted except in
a consensual manner.\\

\noindent \textbf{Security and dependability of Core subsystems: Technology aspects.}\\
In  order to prevent the risks to security and dependability (most
especially abuses against privacy), we have just seen that the policies behind
management and access control of the PriLok Data Vault (PDV) are not single
point. We have explained that \PEA, the group of entities entrusted to manage it,
must be formed following the checks-and-balances principles.

\begin{figure*}[ht]
  \center
  \includegraphics[width=.8\textwidth]{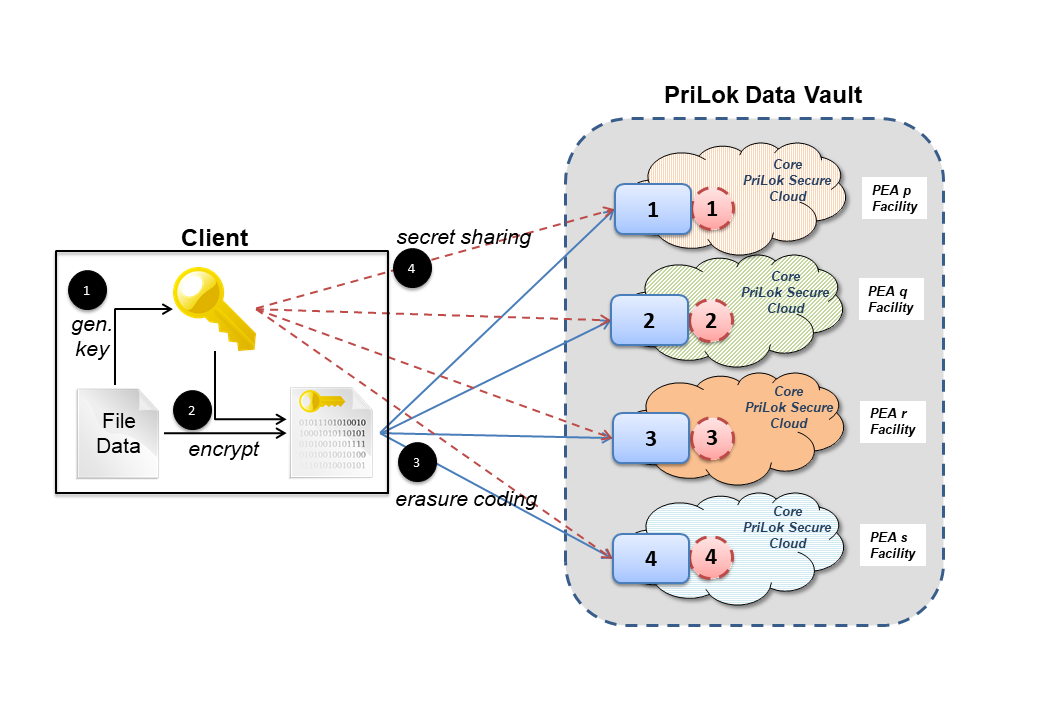}
  \caption{Security and dependability of PriLok Data Vault storage}
  \label{fig:vault}
\end{figure*}

It is important to explain the workings and structure of the PriLok
Data Vault (PDV) construction with a bit more detail, which we do in
Figure~\ref{fig:vault}, as that storage repository assumes an
enormous criticality in the operation of PriLok, since it holds primarily PII.

Again, principles
of distributed fault and intrusion tolerance (a.k.a. Byzantine fault
tolerance, BFT) are followed in the implementation of the mechanisms
controlling the access to the repository, and the repository units
implementing the latter. This middleware transforms the logical
centralisation in physical decentralisation, over a set of distributed
nodes. 
For example, secret
sharing~\cite{sec_sharing} prevents unilateral reconstruction of
confidential information (e.g., by malicious insiders), erasure
coding~\cite{erasure_coding} provides the same property for data
integrity, preventing unilateral damage, and deploying such encoded
data over mutually distrusting clouds~\cite{depsky,scfs} extends these
properties to less trustworthy infrastructures (such as public clouds
or, as is the case for the PriLok Data Vault, private clouds in the
\PEA premises to protect this highly-sensitive data at the highest
degree possible).

The design is based on the works of~\cite{depsky,scfs}. As
Figure~\ref{fig:vault} shows, all starts with a register or file
access request, read or write. Connecting to
Figure~\ref{fig:checks_balances}, this request would be part of the
workflow execution (3), PIS acting as clients.  Let us imagine a write
request. A key is generated on the fly (1), the file encrypted
(2). Then it is split in several pieces by erasure coding (4 in the
example).  Key shares are calculated for the key (4) (4 in the
example). Then, both the file pieces and the key shares are scattered
over several clouds, in several sites. Reading reverses these steps.

Concisely, this design leverages the natural redundancy and
possibility of scattering of PDV over several storage clouds in the
\PEA elements. This has the virtuous effect of complementing the
protection, by reducing the threat surface (the exposure to attacks,
e.g. but not only, on privacy) presented both to external attackers,
and to insiders from within each \PEA member entity. PDV access
through the VPN will thus be controlled by protocols running in the
several core PIS of the \PEA, establishing consensus or matching
thresholds for the operations.

These implementations should be transparent to the users, to preserve the
benefits of logical centralisation, and integrate well with the above-mentioned
workflows. As sophisticated as it may be, there is in fact technology emerging
from research over the past few years, available with a high TRL (technology
readiness level) to make this objective a feasible one. Since we foresee that
ALL operations are systematically attested and certified, the integration of BFT
protocols is also an easy means to effect indelible logging and/or blockchain
supported ledgers (many, if not most of the blockchains of late are implemented
based on BFT).

\pjv{
(cite examples of TRL ? BFTSMART in Hyperledger Fabric; Vawlt (Alysson's
company); LeanXcale CPE Ricardo Jimenez-Peris company; Feedzai CPE Paulo
Marques)
}

\comm{
need a paragraph explaining how the picture above is implemented. It could be
inserted below at the right location. : this is pretty much abt how DepSky, SCFS
works, should be easy to extract text from their intros, because here we do not
want detail. Unlike DepSky, SCFS, which don't care abt the correctness of the
client, important to explain that Client request is the result of a prior
threshold operation where we got the consensus of a quorum of \PEA people to
start a transaction on the Vault (we talk abt that below). What is explained in
the figure is the after, how it works in the vault (which is the present
section, to resist attacks to the cloud)
}

\noindent \textbf{Security and dependability of Core subsystems: Data Protection Regulation
aspects.}\\
We foresee that the operations on the PriLok Data Vault (PDV) secure the
principles of data minimization and storage limitation followed by data
protection authorities and the GDPR in  general. After post-processing of data
extracted, all redundant data must be immediately disposed of, and we are
assuming that the remaining data is the one meaningful for the classical
operation (i.e., without PriLok) of the state services such as the NHS, for
example, the identification of infected, or suspected infected subjects.

N.B.- The current baseline architecture minimizes the threat plane and achieves
high resilience, under the premise that in this first version, the most pressing
requirements R1-R4 are fully met in a highly secure way. In essence, extracting
efficiently and in near real-time, information that would end up in standard
systems, e.g., the NHS, albeit in a much more painful, slow and incomplete way.
For example, the identification of infected, or suspected infected subjects.
Other richer (and useful) services --- e.g., with regard to other requirements,
which we certainly endorse --- possible over the PDV information, should follow
the precautionary principle of general law, as well as the purpose limitation
principle of GDPR. S.o.t.a. research has been showing that preventing
re-identification (de-anonymisation) is a quite difficult task, especially when
one has access to additional spatial-temporal events about the subjects,
acquired by OSINT (open-systems intelligence) or other means~\cite{de2013unique,de2015unique,gymrek2013identifying}. As such, extreme
care should be taken in the handling of that information in a more risk prone
way.



With regard to maintaining the resilience level of the in-core workflows under
more sophisticated operations, further research is suggested on the
investigation and verification of algorithms allowing distributed
privacy-preserving workflows over the VPN, for example leveraging s.o.t.a.
partially homomorphic encryption and secure multiparty computation.

\subsection{Information flow}

Information flow will depend on the system state and the operation mode invoked. \\

\noindent
\textbf{System states:}
At any given moment, the system is in one of the following states:
\begin{enumerate}[1.]
	\item \textbf{\emph{Passive}} - the system is working as a pruning passive listener, and keeps
a minimal amount of information, in the form of PDRs, which are encrypted and
written continuously to/from the Edge PriLok Secure Clouds (PSC) (in Telco
Provider cellular Controlling Data Centres).
However, the PDRs are constantly pruned: only a recent history of PDRs is there,
but inaccessible. The Clouds are locked to operations from the VPN (and reading
from the Provider is technically infeasible).
PriLok Data Vault (PDV) is either empty, or locked for reading or writing,
depending on the implementation approach.  The unlocking of both the vault and
the secure clouds is a highly-critical operation, see below.

	\item \textbf{\emph{Alert}} - the system starts to operate to face a potential epidemic, and
the information flow to and through the core starts.
The Edge PriLok Secure Clouds (PSC) (in Telco Provider Cellular Controlling Data
Centres) and the PriLok Data Vault (PDV) are unlocked.
The unlocking of both the vault and the secure clouds is a highly-critical
operation, see below. In this state, the system core may store raw,
pre-processed and post-processed PDRs, always in encrypted form, through the
period of duration of the alert.

	\commint{ \item
	Emergency   (I THINK THIS WILL NOT BE NECESSARY, it was to represent a state
	(calamity, emergency) where some trade-offs may have to be made w.r.t. PII
	protection vs nation protection. BUT this can be represented by operation modes
	– some will only be possible in that case)
	}
\end{enumerate}


\noindent \textbf{Operation modes:}
There are several operation modes of different criticality, defining different
authorisation (access control clearance) criteria for the different entities.
%
The modes are impacted, amongst other factors, by the criticality of information
with regard to privacy. Critical information is any piece of data that has at
least one PII-critical record):
\begin{itemize}
	\item \textbf{\emph{Lock/unlock}} – operations which materialise the change of state
				from Passive to Alert, or vice-versa, namely and respectively, unlocking
				or locking the core Vault and the edge Secure Clouds, and starting other
				services such as the CEP Engine.
	\item \textbf{\emph{Strict push}} – operations are write-only, no read possible.
	\item \textbf{\emph{Blind analysis}} – operations can read from encrypted data
				(e.g., encrypted searches), and will be supplied the needed metadata.
	\item \textbf{\emph{Blind processing}} – operations can read/write from/to encrypted
				data (e.g., encrypted searches, partially homomorphic update actions),
				and will be supplied the needed metadata.
	\item \textbf{\emph{Full processing}} – operations can read/write from/to cleartext
				data (e.g., record searches, update and record creation actions),
				and will be supplied needed metadata, such as decryption/encryption
				keys.
\end{itemize}

Whenever possible, operations on cleartext data should be done under protection
of Trusted Execution Environments (TEE, such as Intel SGX, or ARM TrustZone).
Information containing critical data should be encrypted before written into a
PriLok repository (e.g., the PriLok Data Vault (PDV)).\\

\noindent \textbf{Notation:}
\begin{description}
  \item [$e^i_p$] - event of ordinal $i$, happening at $p$.
  \item [$t_x(a)$] - real time instant related to $x$, for example $x=0$ - start instant, or [$x=in$] - entry, happening at participant $a$.
\item [$\delta t_x$] - real time interval related to $x$.
\item [$T_x$] - predefined time instant or interval, related to $x$. For example a
message delay, or a timestamp of a real time instant, or other system constant
or variable.
\item [$T^p_{x}$] - idem, happening at participant $p$.
\end{description}

\commint{
MAYBE THIS IS THE POINT WHERE WE INSERT THE INFORFLOW/WORKFLOW figure, AND the
bulleted list of what happens, that we describe next.
}
\ \\

\noindent \textbf{Edge realm:}

\begin{center}
\begin{figure*}[ht]
  \centering
  \includegraphics[width=\textwidth]{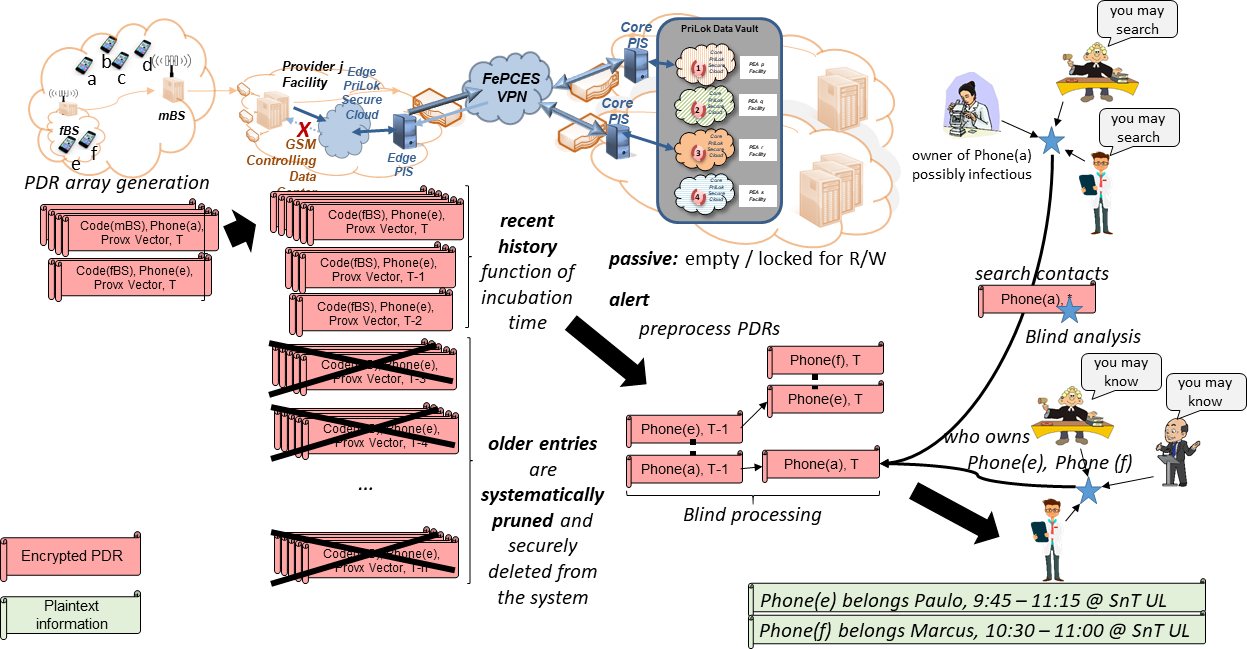}
  \caption{Information flow through the PriLok architecture. The
    recent history of encrypted PDRs is collected in the Edge PriLok
    secure cloud infrastructure and, after the system is alerted,
    blindly preprocessed. Given authority through the federation, PDR
    contacts of positive tested individuals can be searched through
    blind analysis. Once records of possibly infected individuals are found and extracted, their 
identity can be selectively released (e.g., to NHS), given consensual approval through the \FePCES members.
    \comm{We know some old BS cannot encrypt the paging message and
      encryption of the information which goes into PDRs may not be
      desirable by telco providers. However, more modern BS will be
      able to protect this communication and anyway the time this
      information lives until it gets encrypted and forwarded is quite
      short. We therefore consider this a minor security risk in the
      first version of our approach.}
  }
  \label{fig:information_flow}
\end{figure*}
\end{center}

\comm{On second thought, an alternative, maybe more aggressive wrt privacy but
simplifies pre-processing:
\emph{\textbf{APPROACH 3: (generates one PDR array per minute per
all phones inside a BS at that time)}}
}
\noindent \emph{Proximity Detail Record (PDR)} are the source records containing raw
relative proximity data of phones w.r.t. a base station $BS_i$. They are issued
by each base station for each phone $k$ in its area, every minute $p$ of the clock, and thus synchronised at
all providers. They are organised in sets of tuples of the following format:
\begin{equation*}
  \texttt{PDR}^{p}_{k}\left(\texttt{code}(BS_i),~\texttt{Phone}_k(nr,imei), ~\texttt{ProxVector}_k(BS_i), ~T^{p}_{pdr}\right)
\end{equation*}
\begin{equation*}
  \texttt{PDR}^p = \left\{ \texttt{PDR}^{p}_{1}, \cdots, \texttt{PDR}^{p}_{n}\right\}
\end{equation*}

PDRs for a phone $k$ contain the encoded ID of the region where the
phone is (base station $BS_i$), the phone NR and IMEI, the average
proximity vector from the $BS_i$ centroid (a relative polar
coordinate), and the timestamp of creation of that PDR by the $BS_i$ clock,
$T^{p}_{pdr}$). 
All the PDRs from a given time (the same $p$ ``minute'' as above) are organised
in the base station as a set $\texttt{PDR}^p$, which is then encrypted.\\

\comm{
\noindent \emph{\textbf{APPROACH 2: (generates one PDR per minute, per phone inside a BS
at that time)}}
}
\comm{
\emph{Proximity Detail Record (PDR)} are the source records containing raw relative
proximity data w.r.t. a base station $BS_i$. They are issued by each base
station every minute $p$ of the clock, and thus synchronised at all providers.
They are tuples of the following format:}
\comm{
$PDR^{p}_{k}(code(BS_i),Phone(nr,imei),ProxVector(BS_i),T^{p}_{pdr})$
}
\comm{
\noindent \emph{\textbf{APPROACH 1: (generates one PDR per interval inside a BS, per
phone; then I thought abt APPR 2 above, and then APPR 3 up above)}}
}
\comm{
\emph{Proximity Detail Record (PDR)} are the source records containing raw
relative proximity data w.r.t. a base station $BS_i$, of phone $k$. They are
tuples of the following format:}
\comm{
$PDR_k(code(BS_i), Phone_k(nr,imei), T_{in}(BS_i), T_{out}(BS_i),
ProxVector_k(BS_i), T_{pdr})$\\
}
\comm{
PDRs contain, per region (cell $BS_i$): phone NR and IMEI, encoded ID of the
region $BS_i$, the in and out timestamps of contiguous periods of time spent in
the region by the phone, the average proximity vector  from the $BS_i$ centroid
(a relative polar coordinate), and timestamp of creation of the PDR by the
$BS_i$ clock, $T_{pdr}$). In the future, with multi-DSA (dynamic steering
antennas), this info may get even more fine-grain, thus precise.
$T_{in}(BS_i),T_{out}(BS_i)$ – they mark a period from: the first hit on that
cell, to the timeout of the first non-responding paging.}
\comm{
\textbf{Additional general details we should care about:}
}
\comm{
The timestamp of creation of the PDR $T_{pdr}$ is copied as cleartext metadata,
for pruning. Any vulnerability?\\ $T_paging$ period to be short enough to
accompany dynamics (walking speed (e.g., fast is 6~$km/h$), that's how you
infect or get infected) and get a tight enough timeout for $T_{out}$.}
\comm{
More than one macrocell from the same provider may hit the same phone (they
partially overlap) in a small interval. Similarly for picocells in a same paging
cell. This may be used later for precision enhancement (PDR pre-processing).
}
\comm{
There may be a succession of such near positioned PDRs (e.g., subject wandering
inside a shopping mall), staggered in time and even crossing several picocells
(pBS), or PriLok femtocells (fBS).
}
\comm{
The useful diameter and the coordinates of all the BS (absolute and radiation
diagram centroid) are known by the respective providers (and probably
regulators). They also know the hierarchies of all their paging cells. The
coordinates of the PriLok fBS can be obtained once the provider BS coordinates
known. This means that all these coordinates and diameters can also be acquired
by a motivated attacker. As such, the coding of the BS ID code($BS_i$), should
be robust, and/or not accessible to encrypted search, to maintain the relative
proximity nature of the PDRs at rest (and not absolute location).
}
\comm{
The question of obscuring the Phone(nr,imei) to a non-authorised entity doing
encrypted search on a PDR repository is relevant: it is the only way for
authorised entities to locate the PDRs relative to a given phone. But this may
reveal, by OSINT or other, the name/address of the person.}
\comm{
Would be good to estimate the size of PDRs generated per day (in ~20 useful
hours), in a country of 10Mio inhabitants (has +- 250 mBS). Then we could scale.
With the first approach maybe one PDR per 5min per phone. With the second, one
per minute, for all phones in a BS.}

PDRs from
Telco Provider Cellular network cells, macrocells and picocells (and PriLok
femtocells)
are continuously collected by each Provider, encrypted with an asymmetric public
key made available by the \PEAnm (\PEA) to the Providers (or
each provider for fault independence), and then stored in push mode (unilateral
write-only mode), in the 
Edge PriLok Secure Cloud (PSC) (in Telco Provider Cellular Controlling Data
Centres).

The timestamp of creation of the PDR, $T^{p}_{pdr}$ is also annexed as cleartext
metadata, for pruning.
\comm{
(Ok, we discussed a smarter way, and actually, that could be one of the terms
for the encrypted search, since now we store PDRs by the exact clock time,
9:00, 9:01, etc.. Chge)
}
From now on, this data stays at rest and can only be accessed from the
\FePCESnm (\FePCES) VPN, 
through the
\EPISnm (\EPIS)  
(containing the edge services and connection to the VPN).
A time-to-live parameter PDRttl is set to a value defined by the NHS experts.
The rationale is “how long back should tracing go, when a first infection notice
is known?” (this could be “in the country” or, experience advises, “in the
world”). This time will be a function of the disease incubation time,
$T_{incub}$. A value like at least twice or thrice the incubation time gives an
idea. The PIS controls the time-to-live parameter of each PDR record, from its
$T_{pdr}$ and everyday erases, through secure delete, the PDRs whose life has
expired. \\

\noindent \textbf{VPN realm:}
The flow of \emph{critical information} should only be made through the
\FePCESnm (\FePCES) VPN, which runs amongst
the Edge and Core PIS, offering protocols protecting security and dependability
of communication.\\

\noindent \textbf{Core realm:}
Most of the time (hopefully), the system is in Passive state. As such, it is
almost empty of information, as seen above, and both the vault and the secure
clouds are locked.

So, now let us analyse the information flow when the system goes to Alert state,
after being unlocked by a highly-critical operation, see below. In this state,
the system core starts analysing and processing essentially three kinds of
records:
\begin{itemize}
	\item \emph{Raw PDRs} – start coming from all Providers, during the Alert
				interval, as necessary for the workflows.
	\item \emph{Pre-processed PDRs} – Results of analysis of raw PDRs, destined
				to improve the precision of determination of the PDR parameters, as well
				as finding and scoring simple proximity suspicions between pairs of
				phones in the space-time, across different providers.
	\item \emph{Post-processed PDRs} – Results of analysis of pre-processed and/or
				raw PDRs, destined to create insights about infection propagation and
				chains thereof.\\
\end{itemize}

\noindent \textbf{Pre-processing:}

\comm{
This could be done only in reactive mode, when we have a suspected
infected/infectious phone. But might require too much time, even if done in
cleartext, and degrade near real-time response. Preventive pairing to create
suspicions seems better. If could be done in encrypted search the way we want,
would not risk at all. In what follows, phone of interest may be: an alert
(reactive); the systematic spanning of all phones in the repository, until all
pairings found. For ease, we use the approach of the $\mathcal{PDR}^p$ set.
}
The operations below are triggered in consequence of a certified request from
the \PEA, to find out about one or more phone(s) of interest for holders being
(or suspected as being) either infected or infectious (checked with the NHS).
Information given is $\texttt{Phone}_v(nr,imei)$ and estimated earliest infection instant
$\texttt{Phone}_v({T\mathit{inf}}_{min})$, when it is believed that holder was potentially
contaminated.\\

\noindent \textbf{Finding suspicions:}
\label{sec:finding_suspicions}
We start by searching for a phone of interest (and repeated for all
phones of interest). This operation can happen at any time during the
Alert state, so we should narrow the search in function of the
incubation time $T_{incub}$ (this number, supplied by the NHS,
accounts for the worst-case (longer) estimate for the current infection, including the margin of error).

Note that we wish to know both who could have infected the holder of $v$, and
who $v$ could have been infecting after being infected. Given a phone of
interest $\texttt{Phone}_v(nr,imei)$, this means finding all the
$\texttt{PDR}^p = \left\{ \texttt{PDR}^{p}_{1}, \cdots, \texttt{PDR}^{p}_{n}\right\}$ sets where $v$ exists, and such that for all $p$, $T^{p}_{pdr} \ge {T\mathit{inf}}_{min}$.\\

\comm{ NB PJV: I don't remember how i cooked this condition above:
  ${T\mathit{inf}}_{min}-T_{incub}$. Today, it maked no sense to
  me. So i thought better, changed it, here's rationale below. PLEASE VERIFY!\\
How Tinf_min is estimated: 
- patient with symptoms at time Ti : goes back to Ti-Tincub-epsilon, epsilon is error
margin determined by doctor (e.g. he has been like that for say 3 days
before coming to hospital)
- patient without symptoms tested positive at time Ti : find Tinf_min
from estimate of contact date from who contaminated him (if he came
test because of that); or (random test), goes back to
Ti-Tincub-epsilon, assuming worst case that he is at the end of
incubation time.
- corner case: patient without symptoms random tested positive at time
Ti, but who is assymptomatic. Tinf_min could go back quoite a deal but
we cannot know.\\
Then, in the equation above, it only matters the period AFTER he was
infected (w/ margin of error ) it's really only
${T\mathit{inf}}_{min}$. 

}

So, in a time series $p_1, \cdots, p_k$, we have a varying list of phones that have
appeared near $v$, during different parts of the time series. Consider $P$ the
set of all these phones.

\commint{
PROBLEM: We wanted to keep location relative (and not absolute) for as late as
possible, and we did that so far for a single provider. Now as explained
earlier, we have the data from all providers mixed in the Vault, of course. To
make sense of proximity between phones of different providers, we have to
establish a relation between the respective coordinates. Easy way: at this
point, we de-anonymize the BS codes, and obtain their coordinates from the
Providers (was foreseen anyway, for later, in special operations e.g., to find
hot spots). So, to move on, I assume in the below that I have ALL the PDRs from
all the providers, and all references are valid. To see later the best way to do
that.
}

Now, for each pair of phones $(v,u)$, where $u \in {P}$, we find all occurrences that
situate both in some same space-time region (one or several subseries
$p_i, \cdots, p_j$ within the $p_1, \cdots, p_k$ interval).

Inside that group of registers for $(v,u)$, we refine the precision of the notion
of distance (remember we may have events from mBS, pBS, fBS) between the pair,
as well as the notion of duration (remember that there may be noise, and/or both
phones e.g. may wander at a short distance, but between pBS and/or fBS e.g. in a
mall, in an interval of minutes).

Now consider a threshold, to be defined by NHS scientists and technicians, for
the minimum spatio-temporal contact values to raise a potential contamination
suspicion (boolean $PC\mathit{susp}$), of ${Prox}_{max}$ and ${Dur}_{min}$.

\comm{
these criteria exist and have already been used in some countries performing
contact tracing, it is a question of finding out more and refine). (Less
than 5 meters fair, less than 2 meters high, during at least 10 or 15 minutes)
}
We analyse the data, and in result, identify hits of a condition:\\
$PC\mathit{susp}(v,u) = True~\texttt{if}~ (Prox(v,u) \le {Prox}_{max}) \wedge (Dur(v,u)
\ge {Dur}_{min})$

\comm{
(Math has to be refined. Essentially: there is at least one period of at least
${Dur}_{min}$ where they have been at or nearer than ${Prox}_{max}$).
}

\comm{
(At this point, we have suspicions, some may be weak. The function is quite
simple, and maybe it's the right way, we complicate in the next step: In
$PC\mathit{scor}$, we clearly need the integral of the time two phones spent at a given
distance (volume of microbes….). So, MV: «If I sneeze in your face, a < 1sec
contact suffices». OK the example is the extreme, we may not get it, else we
get a stupid FalsePos rate. These are corner cases. But maybe it is worthwhile
mentioning here, so that we set the min/maxes above so that we do not miss too many
likely occasions. Because we will refine in $PC\mathit{scor}$.)
}

After this analysis, we obtain a list of “suspected” potential contamination
pairs.
Now, this list is important for the \PEA entities as a quick though coarse output in reaction
to some event. However, it would be necessary to continue and refine those
suspicions, according to a scale of risk of infection. \\

\noindent \textbf{Scoring suspicions:} We now continue refining the
suspicions. We need to define the confidence we put in each suspicion,
i.e. compute a suspicion score $PC\mathit{scor}$ for each suspected
pair $(v,u)$. Note that there may be input from several suspicion
events, most possibly within a paging cell (mBS, pBS, fBS).  For
simplicity, and without loss of generality, let us call the target of
our scoring effort, a space-time region $R_b$, that is, a certain
interval of time in a certain limited perimeter of space.

This is a multivariable calculation, where heuristics also find a place, in more
refined future versions, especially as the infection mechanisms of the disease
start to be better known. It must be remembered that PriLok is a generic
system, for any epidemic infection to come, possibly unknown. To give it a
start, we define a simple enough function for now:

\begin{equation*}
\begin{split}
PC\mathit{scor}(v,u,R_b) = f( & R_b,~{Prox}_{avg}(v,u), ~{Dur}_{tot}(v,u), ~precision(Prox), \\
 & ~precision(Dur), ~density(R_b), ~severity(R_b))
\end{split}
\end{equation*}

And we create corresponding tuples with both the score and the
function terms, which are stored. This way, scores can be used readily
in first analyses looking e.g. for all the PDRs relating $v$ and $u$,
whereas more sophisticated workflows (under the due authorisations)
can go back to the terms' detail:

\begin{equation*}
\begin{split}
PC\mathit{scor}^{R_b}_{v,u}~( & PC\mathit{scor}(v,u,R_b),R_b,~ {Prox}_{avg}(v,u),~ {Dur}_{tot}(v,u), \\
& precision(Prox), ~precision(Dur), ~density(R_b), ~severity(R_b)) \\
\end{split}
\end{equation*}

\comm{(Math must be refined.)}  $R_b$ is described by the envelope
interval of time of the evaluation and the envelope of the space area
considered (i.e. paging cells). ${Prox}_{avg}(v,u)$,
${Dur}_{tot}(v,u)$, account for the periods of at least ${Dur}_{min}$
where $(v,u)$ have been at or nearer than ${Prox}_{max}$, summing-up
and integrating that time $({Dur}_{tot}(v,u))$, and also considering
how actually close they were on average (even below the max)
(${Prox}_{avg}(v,u)$). That is, if there were 4-5 of periods, e.g.,
walking in a mall separated by short intervals, and they were even
closer than the max, the whole summed-up duration and the real
distance should be reflected in the score. Conversely, if two subjects
were located as being not too near under a same fBS, but they were e.g., sitting in
a restaurant for over a couple of hours, that should as well be reflected
in the score. 

The parameters $precision(Prox)$, $precision(Dur)$, are heuristic
contact evaluation factors accounting for the coverage of the translation of
\emph{digital proximity to actual contact}.
Remember that PriLok assumes a default baseline measurement approach based on
the cellular apparatus, for inclusion, fairness and completeness.
As said before, it welcomes integration of other approaches on a voluntary basis,
which may complement information in specific situations and
areas, e.g., those implemented by GPS, Bluetooth, Wifi or other. 
However, given recent discussions\footnote{https://www.letemps.ch/economie/singapour-tracage-app-degenere-surveillance-masse} care must be taken to make that integration in a way taking into account the non-functional objectives (R7-R10), in
particular digital sovereignty.

Parameter $precision(Prox)$ accounts for a scale of quality of
the method of measurement of distance (mBS, pBS, fBS, BLE, Wifi, GPS, etc.).

Parameter $precision(Dur)$ accounts for a scale of quality of the measurement
of the real infecting contact. It may assume a default value for lack of more
information, but may take into account  specific additional information when the
algorithm is improved, such as speed of trajectory of $v,u$ in $R_s$,
outside/inside, vehicle, stopped (e.g. sleeping), short (at a room), etc.

Parameter $density(R_b)$ is specific of the space-time region and accounts for
the average density of phones (number over useful range) registered in it during
the interval in appreciation. It may be a provider-supplied parameter, or can be
obtained from the PDR data, but can assume a default value for a start. 
Parameter $severity(R_b)$ is again a heuristic parameter that may assume a
default value for lack of more information, but may take into account specific
additional information when the algorithm is improved, e.g., the social role of
$R_s$ area: street, theatre, mall, restaurant, hospital, retirement home,
etc.).

Whatever the function, $PC\mathit{scor}$ will be discretised to assume a range of
discrete values, for practical utilisation by the \PEA. Let us assume a range of
1-4, where highest means highest risk of the potential contamination (this is
conveyed quite well by the function terms, since risk magnitude = probability *
impact): 1- Low; 2- Moderate; 3- High; 4- Very High.

The mission of PriLok in this case is to evaluate the risk of contamination
between a pair of phone holders as precisely as possible, also with input from
\PEA, e.g., w.r.t. to the heuristic parameters. At this point, the diagnostic for
a set of $(v,u)$ pairs is done, both in terms of boolean early warning
suspicions, and a grading of those suspicions. The score gives an opportunity
for selective handling. Several differentiated actions can be triggered as a
function of the score, to be defined by the NHS/\PEA.\\

\comm{
But we can think of a choice of: SMS alert message; phone calls of remote
medicine; person should isolate immediately and be tested at home; come to a
hospital; come to a testing centre; clarify contacts from a list; or just keep
these phones in a cache for follow-up.)
}

\noindent \textbf{Post-processing:}

There will be several avenues for post-processing. Upon analysis of the
pre-processed data, the \PEA entities will decide for several courses of action
w.r.t. each pair, depending on the above risk classes. These may imply further
analysis of the information by PriLok.

An obvious C.o.A. (course of action) for
high enough PC scores of given pairs (as considered by the NHS), besides any
other actions, is to complete the potential contamination findings related to
this pair, by repeating the pre-processing steps for the other phones.

Another obvious C.o.A., as high or very high PC score pairs turn into
infection-positive, besides any other actions, are: find the potential infection
chain (e.g., ordered chains of holders of phone pairs upstream and downstream
some target phones pair); find potential hotspots or infection trajectories
(e.g., resp., a very packed restaurant in fashion, or a bus with one or more
infected persons, riding from a high-level infection area to a remote yet
uninfected town).

We assume again that these operations below are triggered in consequence of a
certified request from the \PEA, to find out about potential infection chains or
potential hotspots or infection trajectories, related to one or more phone(s) of
interest for holders phones having a sufficiently high potential contamination
score (checked with the NHS).\\

\noindent \textbf{Complete potential contamination findings:} 
Given $v,u$ pairs with high enough PC scores, we should re-invoke the
pre-processing steps as above for each $u$, and find all possible $PC\mathit{susp}$
and then $PC\mathit{scor}$ with phones other than $v$.\\

\noindent \textbf{Finding the potential infection chain:} 
In time, the majority of people part of the contacts found in this batch should
have been tested and/or signaled as sick by the NHS. We assume earliest
infection dates ${T\mathit{inf}}_{min}(v)$ were calculated for all $v$.

We go to the repository of pre-processed $PC\mathit{scor}^{R_b}_{v,u}$
registers and create a database containing a new set of registers, $PC\mathit{cont}^{R_b}_{v,u}$
containing only those where $v$ and $u$ are both known contaminated at
current time, and add the respective earliest infection times. We add
as well the coordinates of space-time region $R_b$ where the contact
was identified, as well as the median of the contact interval:
\begin{equation*}
PC\mathit{cont}^{R_b}_{v,u} \left(v, ~u, ~R_b, ~coord(R_b), ~median(\Delta
T_{contact}), ~{T\mathit{inf}}_{min}(v), ~{T\mathit{inf}}_{min}(u) \right)
\end{equation*}

\comm{
(NB: these operations are critical since they reveal absolute locations and
times, and will reveal trajectories)
}

Note that these $PC\mathit{cont}$ registers are annotated versions of the
$PC\mathit{scor}$ registers, they tell the whole history since recorded, and
${T\mathit{inf}}_{min}()$ is added now. So they may refer to contact space-time points
where neither or one of $v$ or $u$ had yet been identified contaminated i.e., it
could be that $\Delta T_{contact} > {T\mathit{inf}}_{min}()$. So, at this point, just by
looking at one register, we do not know whether $v$ infected $u$ or vice-versa.
To find the chain, we have to be able to trace the potential causality in the
real-time domain, between the contact events $PC\mathit{cont}$.

The problem can be reduced to a potential causality determination problem,
leading to a partially ordered directed acyclic graph (DAG), from which many of
the insights desired can be withdrawn. We will rely on a generalisation of
Lamport's ‘happened before' theory for logical channels, to models allowing the
determination of potential causality in the temporal domain for any channels. We
consider the combined analysis of the time-like separation of contamination
events, with the minimum and maximum incubation times as granularity parameters,
and the space-like separation of related contact points between two phone
holders. \\

\comm{
Clues. Given a node of the graph
$PC\mathit{cont}^{R_b}_{v,u}(v,u,R_b,coord(R_b)$,
${T\mathit{inf}}_{min}(v)$,
${T\mathit{inf}}_{min}(u))$,
we should try to confirm the estimates of each ${T\mathit{inf}}_{min}$ versus $\Delta
T_{contact}$. Factor in minimum and maximum incubation times. For $v$ to have
been infected as early as ${T\mathit{inf}}_{min}(v)$, it would have to have been
contaminated the latest by ${T\mathit{inf}}_{min}(v)-Tincub_{min}$.\\ 
}

\noindent \textbf{Finding potential hotspots or infection
  trajectories:} \commint{(For example, the DAG will allow to
  determine with very reasonable certainty who infected who, and
  where, should that be necessary, but we should be careful with this
  info.)}  Note that this will be an evolving process, which will be
updated as more phones from holders tested positive are inserted. This
way we can follow, and at a certain point predict, the trajectories
and evolution of the epidemic.  For example, from the DAG one can
create a georeferenced projection of infection charts: density,
propagation trajectories, etc.  For finding hotspots, we should search
the repository of positive pairs and do a density map according to the
coordinates of the respective space-time regions $R_b$.  As the
epidemic evolves, these tools will allow the NHS/\PEA to make
predictions and decisions quickly, effectively and as accurately and
minimally disturbing as possible.

\comm{\subsection{Elements about mechanisms and protocols}}

\comm{\noindent \textbf{PriLok femtocell dummy base station:}\\}
\comm{(name!) Notes about design of PriLok (fBS), initial JD/MV notes how does this
integrate now. Not too much detail necessary now.}

\comm{\noindent \textbf{Integrating other contact tracing approaches:}\\}
\comm{Would be good to show inclusive to e.g. DP3T … or others to come … don't know
how yet. Why don't phones send fBS their pseudonym IDs a la DP3T ?\\
- voluntary communication of contacts (including but not being limited to the
BLE approaches; actually, DP3T is a bit of a mess: we have guys calling NHS
saying they were in contact with I don't know who who was infected; is there a
timestamp? And a location? That would help, because DP3T (I hope) gives a score
of proximity).}

\jd{the fBS could behave like a phone in D3PT, i.e., generate and exchange ephIDs with others. If someone is sick, our system would identify the ephIDs that were colocated with that person, and inform the D3PT like systems with those IDs. However, this system can only be as good as D3PT is in terms of privacy and robustness.}

\comm{\noindent \textbf{Consensual PriLok operations:}\\}
\comm{Give outline and discuss how consensus is reached amongst a quorum of elements
of \PEA, in the presence of flts and attacks. Briefly discuss:  the need for k+1
threshold libraries (examples?) for threshold authorisation signatures, and 
secret-sharing for key recovery (PDRs); BFT-SMR or BFT Quorum libraries for
consensus (BFTSMART, etc.), for macro transactions, such as a set of operations
submitted to the depsky client below (ACTUALLY, as I write, the depsly7SCFS
client below, could be expecting f+1 valid requests before proceeding…).}

\comm{\noindent \textbf{Core PriLok Secure Clouds and Vault:}}
\comm{Give outline and discuss org of Core PriLok Secure Clouds (PSC) (in the \PEA
Data Centres), as DepSky like cloud-of-clouds, one partition per entity site,
read/update prot runs amongst Core PriLok Information Switches (Core PIS).}

\comm{
\begin{figure*}[ht]
  \begin{center}
  \includegraphics[width=.7\textwidth]{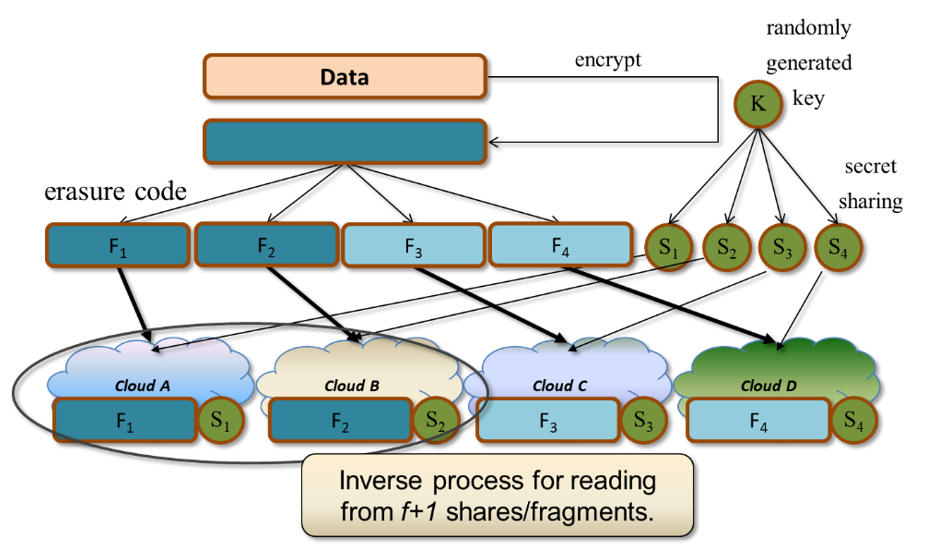}
  \caption{PriLok Secure Clouds}
  \label{fig:clouds}
  \end{center}
\end{figure*}
}

\comm{\noindent \textbf{Searchable and partially homomorphic encryption:}}
\comm{MAYBE later, depending on how we advance, the idea here would be to show how far
we go into releasing as little info as possible in the pre-proc and post-proc
phases in the core, over the Vault:\\
- search encrypted phone ids, or $B_s$ ids, or time intervals, etc.\\
- construct modified registers by adding material from 2 or more separate
encrypted registers - etc.\\
(Notes: 
- searchable encryption – find a register e.g. phone id , in an sql database –
typ hundreds msec. to sec.}

\commint{
- recent schemes of searchable encryption are based on assym encryption, to
minimize leakage. FIND HOW/WHICH. This is what we need (encrypt all PDR in
periphery with same pub key, or even one by provider).\\
- then the priv key was shared and deleted. (we may need proactive secret
sharing not to lose the key!). And then all these critical decryption ops over
the Vault (like stealing PII) cannot be made by one rogue guy, ext or inside one
entity, takes k+1 rogue guys!:)\\
- back to searches: they are are actually made, not over the encrypted
data, but over an encrypted index that is fabricated when the encrypted DB is
made. So (a la large-scale file systems) the index metadata can even be kept in
a site different from the data!, with diff security properties.\\
- Now, this is great, we can also protect the index, for not letting just anyone
go find the encrypted regs that contain a phone number, remember?? (i.e.
constraining access to the metadata to authorised ops.).\\
- Also, we can have selective/imbricated encryption in the registers (edge PDR
or core PC), and we just have to create different search indices! (remember
that in the pots-proc, or even in pre-proc, we will have to have coord and
time which are very sensitive PII).\\
- Then, for active ops on registers, given that the stuff may not work well with
homomorphisms in a pure asym crypto, perhaps we might then leverage that we
can encrypt data and index differently to allow some homomorphic crypto ops,
don't know…\\
NB: According to Bernardo Ferreira (an expert at FCUL), all this is new ideas,
so careful, it may be one part of PriLok that leads to research!}

\comm{- Also, for ops on registers, given that the stuff is in the cloud-of-clouds
that compose the Vault, if it does not work well with homomorphisms in this
asym crypto, we might have a look at Secure Outsourced Computation methods,
SMPC, etc.
\url{https://www.morganclaypoolpublishers.com/catalog_Orig/samples/9781627057929_sample.pdf}
}

\comm{\noindent \textbf{ML/AI assisted infection chain and trajectory analysis}\\}
\comm{MAYBE later, depending on how we advance, OSINT ML/AI to improve accuracy,
(suggest sources)}
\commint{What More in SECTION ON PROTOCOLS?}
\comm{
\begin{itemize}
	\item Core PriLok Information Switches (Core PIS).
	\item Core PriLok Secure Clouds (PSC) (in the \PEA Data Centres).
	\item PriLok Complex Event Processing Engine (PCEPE), in \PEA Data Centres.
\end{itemize}
}

\comm{PriLok Data Vault  (PDV), the logically centralised repository.  Although
logically centralised, the PriLok Data Vault  (PDV) construction is NOT
physically centralised.  PDV is a data store, probably key-value, implemented by
one or several private storage clouds. Queries, and direct reads and writes made
on PDV, by quorums of the \PEA entities.}
\comm{
PriLok follows generically a 2q-eyes access control policy. (x+1, n) multiparty
crypto operation,  recovery of (x+1,n) shared key, f+1-fault tolerant quorum.
}
\comm{
Request by one of the entities of \FePCES, authorised by enough other entrusted
entities.  Streamlined and based on IT-supported workflows.
}
\comm{
PriLok Data Vault  (PDV) scattered over several storage clouds in the \PEA
elements.  PDV access through the VPN, by protocols running in the several core
PIS of the \PEA.
}
\comm{
After post-processing of data extracted, all redundant data must be immediately
disposed of.
}
\comm{
Remaining data has lower risk than initial data, since it is of the type of NHS
classical operation.
}

\comm{\noindent \textbf{Edge realm (\APE):} \\}

\comm{There can be import/export of data from/to \APEnm (\APE). 
\APE that only need to receive or send non-critical information do not need
dedicated interfaces. \APE that need to receive or send critical information will
have an
\EPISnm (\EPIS)
installed, to communicate via the \FePCES VPN.}

\subsection{Security, Dependability and Resilience analysis}

Security of information treatment from the base stations down to the core
servers is the responsibility of  the infrastructure holders. The incumbents should
collectively ensure:
\begin{itemize}
	\item Storage of PDRs at the Edge Clouds with multiparty encryption technology.
	\item Minimisation of storage of PDRs at the Edge Clouds, by continued periodical deletion.
	\item Lock of the Edge Clouds for Provider read access.
	\item Full lock of the Edge Clouds during passive state.
	\item Fault and intrusion tolerance of the Core Clouds, by:
	\begin{enumerate}[(i)]
		\item enforcing $k_i+1$ entities to contribute to authorise and/or certify
					in ledger any critical operation;
		\item enforcing $k_j+1$ shares to reconstruct any decryption key;
		\item enforcing $f+1$ diverse nodes to reach consensus on operations or sets
					thereof on the Clouds;
		\item considering quorums of diverse software/hardware replicas to reach
					availability in the face of faults or attacks
		\item enforcing highly secure and robust communication on the VPN.
	\end{enumerate}
	\item Minimisation of operations in the clear on critical data, leveraging:
	\begin{enumerate}[(i)]
		\item utilisation of searchable encryption technology to the extent possible;
		\item minimisation of cleartext manipulation risk by leveraging TEEs in the
					compute clouds.
	\end{enumerate}
	\item Minimisation of critical data storage in the Core Clouds, namely Vault, by: 
	\begin{enumerate}[(i)]
		\item eliminating data as it becomes not needed after being processed, 
					during the Alert state
		\item performing secure delete of all data, as permitted by regulations,
					as soon as the system enters Passive state.
	\end{enumerate}
\end{itemize}

\subsection{Interaction with other societal systems}


PriLok is destined to fulfill several societal objectives and as such
close interaction is expected with these entities.
In particular, PriLok needs to be configured with parameters to be
defined by NHS scientists and technicians, and refined throughout its
operation. For example, when searching for suspicious encounters (see
finding suspicions on page~\pageref{sec:finding_suspicions}), the
incubation time needs to be adjusted to the knowledge epidemiologists
have gathered about the current infection.

Epidemiology experts are also expected to benefit from post-processed
information supplied by PriLok, especially during the active stages
of infections and epidemics.
However, even though PriLok would rely on established procedures and
regulatory frameworks (approval from ethics committees, etc.) to grant
access to this information on an urgent need to know basis, by
enabling authorized entities to extract sanitized statistical
information and pseudomised data sets, we believe further research is
required to ensure the protection of citizens rights, in particular
privacy, for less urgent needs.

\subsection{Cross-border interoperability}


\begin{figure*}[ht]
  \includegraphics[width=\textwidth]{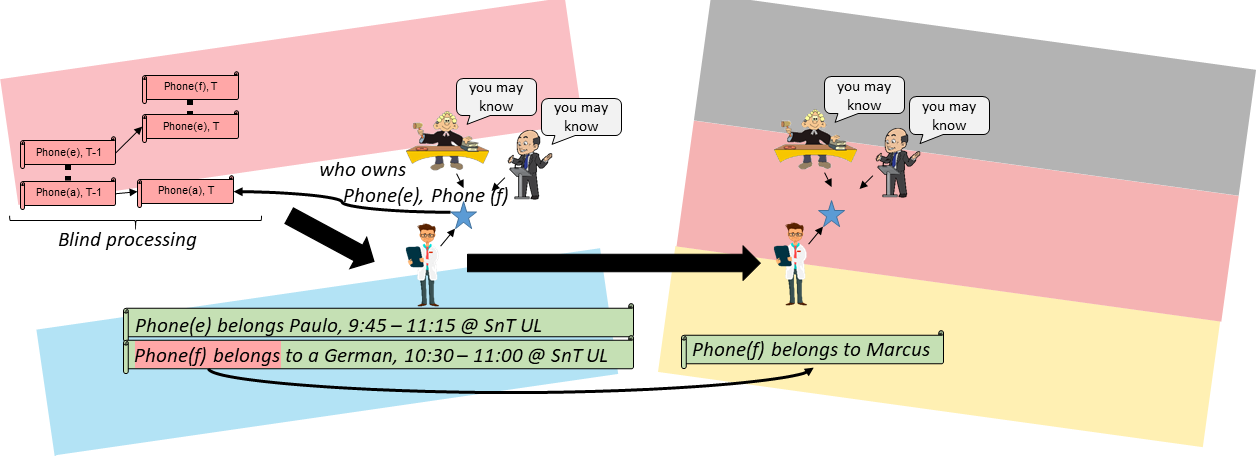}
  \caption{Cross-border infection chain tracing through PriLok}
  \label{fig:cross_border}
\end{figure*}

One aspect of interaction we wish to highlight here, is the
interoperability with systems applied in other countries.  PriLok is
by design a single nation system in the sense that through PriLok
citizens entrust their PII data to the federation of entrusted legal
authorities, either elected, or appointed by an
elected government, and which form the \FePCES. As such, federations
or members of other countries have in principle no right over these citizens' PII data. 

However, it is of course essential to be able to follow infection
chains across countries and to alarm the respective authorities about
the possibility of an infection, or worse a new outbreak.

Much like roaming supports foreigners to obtain access to the mobile
communication network, PriLok is trivially cross-border interoperable
by not revealing the identity of foreigners to another countries
\FePCES. Instead, the final step of reidentifying the person behind
the PDRs it creates is reserved for the country this person lives
in. More specifically, PDRs ultimately can reveal the space-time
coordinates where infections may have happened and the contacts this
person had, including the country she lives in, but to reidentify this
person, authority of the \FePCES of this person's country will be
needed. Figure~\ref{fig:cross_border} illustrates this point.

Barring the technical details, the existing good collaboration between
national health institutes in Europe and world wide, already suffices
to continue tracing infection chains across borders by a simple
exchange of those found encrypted phone-identifying tokens, which only
the \FePCES of the respective country will be able to decrypt to
reveal the person behind. Although much easier with PriLok instances
on both sides of the border, which continuously track the infected and
his contacts through PDRs, the possibility of the home country's
\FePCES to learn about the phone and its owner continues to work with
fundamentally other tracing systems.

\subsection{Acknowledgments}
We could not have created this preliminary design specification
without the help and contributions of a number of people. The fact of
them being experts at different levels of the architecture and of the
hardware/software stacks on which PriLok is built, substantiates our
words in the beginning, that this a complex distributed critical
information systems problem which needs diverse skills such as
distributed algorithms, fault and intrusion tolerance, networking and
cloud technology, cryptography, amongst others, not forgetting the
contribution of the medical fields to establish requirements and needs.
We would particularly like to express our special thanks to
Rui Aguiar, Alysson Bessani, Adam Lackorzynski, and Bernardo
Rodrigues. Thanks for helping out when and where we had doubts or struggled.


\printbibliography

\end{document}